\documentclass[letterpaper,11pt]{article}
\usepackage{amsmath,mathrsfs,amssymb,amsfonts,amsthm,mathtools}
\usepackage{physics,color}
\usepackage{subcaption}
\usepackage{tabularx}
\usepackage{multirow}
\usepackage{booktabs}
\usepackage[table]{xcolor}
\usepackage{tikz,tikz-cd}
\usetikzlibrary{shapes.geometric, arrows, positioning}
\usepackage{empheq}
\usepackage{bm}

\newcommand*\widefbox[1]{\fbox{\hspace{2em}#1\hspace{2em}}}
\pdfoutput=1 
\usepackage{jheppub} 
\usepackage[T1]{fontenc} 

\newcommand{\nn}{\nonumber}

\hypersetup{
    colorlinks=true,
    linkcolor=blue,
    citecolor=blue,
    urlcolor=blue
}

\title{Modulated Accelerating Mirrors as a Physical Realization of the Kappa-Gamma Vacuum}

\author[a]{Arash Azizi}

\affiliation[a]{{\it The Institute for Quantum Science and Engineering,
Texas A\&M University,\\ College Station, TX 77843, U.S.A.}}

\emailAdd{sazizi@tamu.edu}

\abstract{
Modulated accelerating mirrors provide a concrete dynamical origin for the $\kappa\gamma$ vacuum—a thermal, single-mode squeezed state with a tunable angle. The Carlitz–Willey trajectory fixes the Planckian weights (set by $\kappa$), while a weak, chiral, frequency-diagonal boundary drive—equivalently a time-dependent Robin impedance—rotates the squeeze angle (set by $\gamma$) without changing those weights at leading order. On future null infinity, the two-point function cleanly splits into a stationary thermal piece and a phase-sensitive, non-stationary piece. Inertial Unruh–DeWitt detectors see an exact Planck law; uniformly accelerated detectors expose $\gamma$ through interference and can show mode-selective suppression under frequency matching. Numerical wave-packet simulations corroborate the phase imprint and parametric amplification. In short: \emph{trajectory sets scale, boundary sets angle}. This separation turns abstract squeeze parameters into laboratory-tunable signatures and offers a practical route to engineer and diagnose $\kappa\gamma$ vacua in moving-mirror analogs.
}

\begin{document} 
\maketitle
\flushbottom

\section{Introduction}
\label{sec:introduction}

A cornerstone of modern theoretical physics is the discovery that the quantum vacuum is a dynamic, observer-dependent entity. Phenomena such as the Unruh and Hawking effects have revealed that observers in non-inertial motion or in the presence of event horizons perceive the Minkowski vacuum as a thermal bath of particles \cite{Unruh1976, Hawking1975, Fulling1973, DeWitt1975PhysicsRep, Birrell_Davies1982}. This insight is typically understood from two complementary viewpoints: a kinematic one, in which Bogoliubov transformations relate mode algebras across observers, and a dynamic one, in which time-dependent boundaries or background geometries physically create quanta.

From the dynamic viewpoint, accelerating boundaries in $(1+1)$ dimensions provide an exactly solvable laboratory. Beginning with Moore’s time-dependent cavity field theory \cite{Moore1970} and the Fulling–Davies program \cite{Fulling_Davies1976,Fulling_Davies1977}, particle creation is encoded by a ray-tracing map $v=p(u)$ that relates incoming to outgoing null rays (see also \cite{Birrell_Davies1982}). The exponential Carlitz–Willey trajectory yields a constant thermal flux with temperature set by the acceleration scale \cite{Carlitz_Willey1987}. This framework underpins a broad connection between black-hole evaporation, the dynamical Casimir effect, and analogue experiments \cite{Dodonov2020review,Chen_Mourou2017PRL}, and has been refined in time-resolved and quasi-thermal settings \cite{Good2013mirror,Good2020Wilczek}.

Our kappa program began from a different angle: a kinematic curiosity about Unruh modes \cite{Unruh1976, UnruhWald1984}. Because the Unruh construction is intricate, we asked how to reproduce its structure starting from Rindler modes but allowing a continuous weighting of the coefficients. By replacing the fixed Unruh coefficients with a one-parameter family, we obtained a $\kappa$–Rindler mode that recovers the Unruh mode at $\kappa=1$ (and hence the Minkowski vacuum) and the Rindler vacuum as $\kappa\to 0$ \cite{Azizi2022Kappashort,Azizi2023JHEP}. These states  had direct physical consequences, leading to a tunable Unruh effect where an accelerated detector could perceive a temperature either hotter or colder than the standard Unruh value \cite{Azizi2023JHEP, Azizi2025Tunable}. Following the same mathematical strategy, this construction was adapted to inertial frames, yielding the \textit{$\kappa$-plane wave vacuum}---a thermal squeezed state that smoothly reduces to the Minkowski vacuum as $\kappa\to 0$ \cite{Azizi2025KappaPW}. A further generalization introduced a phase parameter $\gamma$ to define the \textit{$\kappa\gamma$-vacuum}, which incorporates a squeeze angle and gives rise to non-stationary, phase-dependent correlations \cite{Azizi2025KappaGamma}. While these $\kappa$-states were mathematically well-defined, their physical origin was initially an open question.

A physical origin for the purely thermal ($\gamma=0$) member of this family was recently found in the dynamic picture of moving mirrors \cite{Azizi2025Mirror_KappaPW}. Specifically, the quantum state produced by an accelerating Carlitz-Willey (CW) mirror is operationally identical to the $\kappa$-plane wave vacuum on \textit{future null infinity} ($\mathscr I^+$), the asymptotic boundary where radiated fields are measured. This state has the mathematical structure of a single-mode \emph{SU(1,1)} squeeze, and its thermality is formally guaranteed by the Kubo–Martin–Schwinger (KMS) condition \cite{Kubo1957, Martin_Schwinger1959}. The equivalence is exact in this asymptotic region where left- and right-moving modes decouple, allowing for a direct comparison. This discovery provided a crucial physical interpretation, identifying the kinematic parameter $\kappa$ with the acceleration of the CW mirror's trajectory.

The standard CW mirror, however, only accounts for the purely thermal effects associated with $\kappa$. This raises the central question of this work: given that the mirror's \emph{trajectory} sources $\kappa$ on null infinity, what corresponding physical process in the mirror model can source the phase $\gamma$?
This paper answers that question by demonstrating that $\gamma$ is dynamically generated by the mirror's \emph{boundary interaction}. We provide the physical realization for the complete $\kappa\gamma$-vacuum by solving a model of a CW mirror endowed with a \emph{time-dependent reflectivity}, implemented via a Robin boundary condition. We show that the phase of this temporal modulation on the boundary directly controls the phase $\gamma$ of the asymptotic quantum state on future null infinity.

Our proof is comprehensive. We systematize the Bogoliubov transformation for the $\kappa\gamma$-family, derive the corresponding Wightman function, and analyze the response of Unruh-DeWitt detectors. This analysis reveals that $\gamma$ is a physically measurable parameter for accelerated observers. We conclude with numerical simulations of wave packet scattering that visually confirm how a modulated boundary interaction generates the complex, amplified, and phase-shifted field correlations characteristic of the $\kappa\gamma$-vacuum. This work thus completes the physical picture of this family of vacua, grounding an abstract kinematic construction in a concrete, controllable dynamical system.

The paper is organized as follows. Section~\ref{sec:bogol} details the Bogoliubov transformations for the $\kappa\gamma$-family. Section~\ref{sec:wightman-kms} derives the Wightman function and analyzes its KMS structure. Section~\ref{sec:udw_response} calculates the response of Unruh-DeWitt detectors. Section~\ref{sec:dynamical_origin_merged} presents the central result, deriving the modulated mirror model. The appendices contain detailed mathematical derivations.

\section{Bogoliubov transformations}
\label{sec:bogol}

In this section, we spell out the Mellin–diagonal $SU(1,1)$ structure of the $\kappa\gamma$ family at a fixed log–frequency label $\Lambda>0$. We will (i) define the single–mode (in $\Lambda$) Bogoliubov map between two members of the family, (ii) give several equivalent parameterizations, (iii) work out checks and limiting cases, and (iv) recall the moving–mirror realization that reproduces the same single–mode squeeze at $\mathscr I^+$.

We work with the light-cone coordinates $u = t - x$ and $v = t + x$. The right- and left-traveling field operators are constructed by integrating over a basis of modes labeled by the positive Fourier label $\Lambda$:
\begin{align}
    \Phi_{\text{RTW}}(u) &= \int_0^\infty d\Lambda \, \Phi_{\Lambda,\kappa,\gamma}(u) \, {\cal A}_{\Lambda,\kappa,\gamma} + \text{h.c.}, \nn\\
    \Phi_{\text{LTW}}(v) &= \int_0^\infty d\Lambda \, \Phi_{\Lambda,\kappa,\gamma}(v) \, \widetilde{\cal A}_{\Lambda,\kappa,\gamma} + \text{h.c.}, \label{eq:kg_mode_field}
\end{align}
where ``h.c.'' stands for Hermitian conjugate. The convenient mode basis for the $\kappa\gamma$ family is given by
\begin{align}
    \Phi_{\Lambda,\kappa,\gamma}(u)
    = \frac{1}{\sqrt{{\cal N}_{\Lambda,\kappa}}}\Big(
    e^{\frac{\pi\Lambda}{2\kappa}+i\gamma}\,e^{-i\Lambda u}
    +e^{-\frac{\pi\Lambda}{2\kappa}-i\gamma}\,e^{+i\Lambda u}\Big), \label{eq:kg_mode}
\end{align}
with the normalization factor ${\cal N}_{\Lambda,\kappa}=8\pi\Lambda\,\sinh(\pi\Lambda/\kappa)$.

The creation and annihilation operators, ${\cal A}_{\Lambda,\kappa,\gamma}^{\dagger}$ and ${\cal A}_{\Lambda,\kappa,\gamma}$, define the $\kappa\gamma$-vacuum via the condition ${\cal A}_{\Lambda,\kappa,\gamma} \ket{0_{\kappa\gamma}} = 0$. They obey the canonical commutation relation:
\begin{equation}
    [{\cal A}_{\Lambda,\kappa,\gamma},{\cal A}^\dagger_{\Lambda',\kappa,\gamma}]=\delta(\Lambda-\Lambda').
\end{equation}
The operator $\widetilde{\cal A}_{\Lambda,\kappa,\gamma}$ denotes the corresponding annihilation operator for the independent left-moving sector.

\medskip

\subsection{\texorpdfstring{General $(\kappa,\gamma)\!\to\!(\kappa',\gamma')$ map at fixed $\Lambda$}{General (kappa,gamma)->(kappa',gamma') map at fixed Lambda}}
\label{sec:bogol-general}

The Mellin–diagonal Bogoliubov map between two members of the family reads
\begin{empheq}[box=\widefbox]{align}
{\cal A}_{\Lambda,\kappa',\gamma'}
=\alpha_\Lambda\,{\cal A}_{\Lambda,\kappa,\gamma}
+\beta_\Lambda\,{\cal A}^{\dagger}_{\Lambda,\kappa,\gamma}.
\label{eq:kgkg-bogol}
\end{empheq}
With
\[
\Delta\gamma:=\gamma'-\gamma,\qquad
r_\pm:=\frac{\pi\Lambda}{2}\!\left(\frac{1}{\kappa'}\pm\frac{1}{\kappa}\right),
\]
the Klein–Gordon inner products give
\begin{empheq}[box=\widefbox]{align}
\alpha_\Lambda
=\frac{\sinh(r_+-i\Delta\gamma)}
{\sqrt{\sinh(\pi\Lambda/\kappa)\,\sinh(\pi\Lambda/\kappa')}},
\qquad
\beta_\Lambda
=\frac{\sinh(r_- -i\Delta\gamma)}
{\sqrt{\sinh(\pi\Lambda/\kappa)\,\sinh(\pi\Lambda/\kappa')}}.
\label{eq:alphabeta-def}
\end{empheq}
Canonicity follows from $|\sinh(x-i\delta)|^2=\sinh^2x+\sin^2\delta$ and \newline
$\sinh^2 r_+-\sinh^2 r_-=\sinh(\tfrac{\pi\Lambda}{\kappa})\sinh(\tfrac{\pi\Lambda}{\kappa'})$:
\begin{equation}
|\alpha_\Lambda|^2-|\beta_\Lambda|^2=1
\quad\Rightarrow\quad
\big[{\cal A}_{\Lambda,\kappa',\gamma'},{\cal A}^{\dagger}_{\Lambda',\kappa',\gamma'}\big]=\delta(\Lambda-\Lambda').
\label{eq:canonicity}
\end{equation}
The inverse is obtained by swapping primed/unprimed parameters and $\Delta\gamma\to-\Delta\gamma$:
\begin{align}
{\cal A}_{\Lambda,\kappa,\gamma}
=&\alpha_\Lambda^{(\mathrm{inv})}\,{\cal A}_{\Lambda,\kappa',\gamma'}
-\beta_\Lambda^{(\mathrm{inv})}\,{\cal A}^{\dagger}_{\Lambda,\kappa',\gamma'},
\nn\\
\alpha_\Lambda^{(\mathrm{inv})}=&\frac{\sinh(r_+ + i\Delta\gamma)}{\sqrt{\sinh\!\big(\frac{\pi \Lambda}{\kappa}\big)\,\sinh\!\big(\frac{\pi \Lambda}{\kappa'}\big)}},
\quad
\beta_\Lambda^{(\mathrm{inv})}=\frac{\sinh(r_- - i\Delta\gamma)}{\sqrt{\sinh\!\big(\frac{\pi \Lambda}{\kappa}\big)\,\sinh\!\big(\frac{\pi \Lambda}{\kappa'}\big)}}.
\label{eq:inverse}
\end{align}

\paragraph{Single–mode $SU(1,1)$ parametrization.}
Each block can be written as
\begin{align}
\alpha_\Lambda=e^{i\varphi_\Lambda}\cosh r_\Lambda,\qquad
\beta_\Lambda=e^{i(\varphi_\Lambda+\theta_\Lambda)}\sinh r_\Lambda,\qquad r_\Lambda\ge0,
\label{eq:squeeze-param}
\end{align}
with squeeze magnitude and relative phase extracted from \eqref{eq:alphabeta-def}:
\begin{align}
\sinh r_\Lambda
=&|\beta_\Lambda|
=\frac{\sqrt{\sinh^2 r_-+\sin^2\!\Delta\gamma}}
{\sqrt{\sinh(\frac{\pi \Lambda}{\kappa})\,\sinh(\frac{\pi \Lambda}{\kappa'})}},
\nn\\
\theta_\Lambda=&\arg\!\big[\sinh(r_--i\Delta\gamma)\big]-\arg\!\big[\sinh(r_+-i\Delta\gamma)\big].
\label{eq:r-theta-extract}
\end{align}
A global phase gauge ${\cal A}\!\to e^{-i\gamma}{\cal A}$, ${\cal A}'\!\to e^{-i\gamma'}{\cal A}'$ removes $\varphi_\Lambda$, leaving only the relative angle $\Delta\gamma$ in observables.

\paragraph{Composition and operator form.}
Package the map as
\[
{\sf S}(\alpha_\Lambda,\beta_\Lambda):=
\begin{pmatrix}
\alpha_\Lambda & \beta_\Lambda\\
\beta_\Lambda^{*} & \alpha_\Lambda^{*}
\end{pmatrix},
\qquad \det{\sf S}=1.
\]
Two successive changes $(\kappa,\gamma)\!\to\!(\kappa',\gamma')\!\to\!(\kappa'',\gamma'')$ compose as
\begin{align}
\begin{pmatrix}
\alpha'' & \beta''\\ \beta''{}^* & \alpha''{}^*
\end{pmatrix}
=
\begin{pmatrix}
\alpha' & \beta'\\ \beta'{}^* & \alpha'{}^*
\end{pmatrix}
\begin{pmatrix}
\alpha & \beta\\ \beta^* & \alpha^*
\end{pmatrix},
\qquad
\alpha''=\alpha'\alpha+\beta'\beta^*,\quad
\beta''=\alpha'\beta+\beta'\alpha^*.
\label{eq:composition}
\end{align}
When $\kappa'=\kappa$ (no change of thermal scale), $r_-=0$ and the transformation is a pure single–mode angle rotation (relative phases add).

Equivalently, at fixed $\Lambda$,
\begin{align}
\begin{pmatrix}
{\cal A}_{\Lambda,\kappa',\gamma'}\\[2pt]
{\cal A}^{\dagger}_{\Lambda,\kappa',\gamma'}
\end{pmatrix}
=
\begin{pmatrix}
\alpha_\Lambda & \beta_\Lambda\\
\beta_\Lambda^{*} & \alpha_\Lambda^{*}
\end{pmatrix}
\begin{pmatrix}
{\cal A}_{\Lambda,\kappa,\gamma}\\[2pt]
{\cal A}^{\dagger}_{\Lambda,\kappa,\gamma}
\end{pmatrix},
\qquad
|\alpha_\Lambda|^2-|\beta_\Lambda|^2=1,
\label{eq:SU11}
\end{align}
which makes \eqref{eq:composition} manifest and cleanly separates the roles of $\kappa$ (thermal scale) and $\gamma$ (global squeeze angle).

\medskip

\subsection{Checks, number spectra, and special limits}
\label{sec:bogol-limits}

The expected number of $(\kappa',\gamma')$ quanta in $|0_{\kappa\gamma}\rangle$ is
\begin{empheq}[box=\widefbox]{align}
\big\langle 0_{\kappa\gamma}\big|\,{\cal N}^{(\kappa',\gamma')}_\Lambda\,\big|0_{\kappa\gamma}\big\rangle
=|\beta_\Lambda|^2
=\frac{\sinh^2 r_- + \sin^2\!\Delta\gamma}
{\sinh\!\big(\tfrac{\pi \Lambda}{\kappa}\big)\,\sinh\!\big(\tfrac{\pi \Lambda}{\kappa'}\big)}.
\label{eq:number}
\end{empheq}
Hence we find a neat reciprocity: the spectrum depends only on the \emph{relative} $SU(1,1)$ squeeze between the two vacua and is invariant under exchanging the roles of “state’’ and “basis,’’ $(\kappa,\gamma)\leftrightarrow(\kappa',\gamma')$. Indeed, $r_-\to -r_-$ and $\Delta\gamma\to-\Delta\gamma$ leave $\sinh^2 r_-+\sin^2\!\Delta\gamma$ unchanged, while the symmetric denominator $\sinh\!\big(\tfrac{\pi\Lambda}{\kappa}\big)\sinh\!\big(\tfrac{\pi\Lambda}{\kappa'}\big)$ is manifestly invariant, so that
\begin{align}
\boxed{\quad \big\langle 0_{\kappa\gamma}\big|{\cal N}^{(\kappa',\gamma')}_\Lambda\big|0_{\kappa\gamma}\big\rangle
=
\big\langle 0_{\kappa'\gamma'}\big|{\cal N}^{(\kappa,\gamma)}_\Lambda\big|0_{\kappa'\gamma'}\big\rangle. \quad}
\end{align}
It turns out that the numerator cleanly separates “temperature detuning’’ (\,$\sinh^2 r_-$\,) from a pure squeeze–angle mismatch (\,$\sin^2\!\Delta\gamma$\,), while the denominator sets the thermal weights of the two references. Consequently, particle production at a fixed Mellin label $\Lambda$ is a basis–agnostic measure of this mismatch and vanishes \emph{only} when both the detuning and the angle difference are zero, i.e., $\kappa'=\kappa$ and $\Delta\gamma=0$.

\begin{enumerate}
\item \textit{Minkowski reference $\kappa'\to0$.}
Since $r_-\to \frac{\pi\Lambda}{2\kappa'}\to\infty$, the phase term is exponentially subleading, and
\begin{align}
\lim_{\kappa'\to 0}|\beta_\Lambda|^2
=\frac{1}{e^{2\pi\Lambda/\kappa}-1},
\qquad T_\kappa=\frac{\kappa}{2\pi}.
\label{eq:planck}
\end{align}
\item \textit{Equal temperatures, phase shift only $\kappa'=\kappa$.}
Here $r_-=0$ and
\begin{align}
\alpha_\Lambda=\cos\Delta\gamma - i\,\coth\!\Big(\frac{\pi\Lambda}{\kappa}\Big)\sin\Delta\gamma,\quad
\beta_\Lambda=-\,\frac{i\,\sin\Delta\gamma}{\sinh(\frac{\pi\Lambda}{\kappa})},\quad
|\beta_\Lambda|^2=\frac{\sin^2\!\Delta\gamma}{\sinh^2(\frac{\pi\Lambda}{\kappa})},
\label{eq:equal-k}
\end{align}
so $\Delta\gamma$ acts as a global squeeze angle: it rotates the quadratures, but number observables remain governed by the same thermal scale $\kappa$.

\item \textit{No-go for angle removal if $\kappa'\neq\kappa$.}
Because $\beta_\Lambda\propto\sinh(r_- - i\Delta\gamma)$ and $r_->0$, there is no real $\Delta\gamma$ with $\sinh(r_- - i\Delta\gamma)=0$; a pure angle rotation (no particle production) is possible only when $\kappa'=\kappa$.

\item \textit{UV/IR edges.}
As $\Lambda\to\infty$, $\sinh(\pi\Lambda/\kappa)\sim\tfrac{1}{2}e^{\pi\Lambda/\kappa}$ exponentially damps $|\beta_\Lambda|^2$. As $\Lambda\to0^+$, $|\beta_\Lambda|^2$ remains finite; the phase channel is additionally suppressed by $\sin^2\Delta\gamma/\sinh^2(\pi\Lambda/\kappa)$.
\end{enumerate}

\begin{figure}[h!]
\centering
\includegraphics[width=0.7\textwidth]{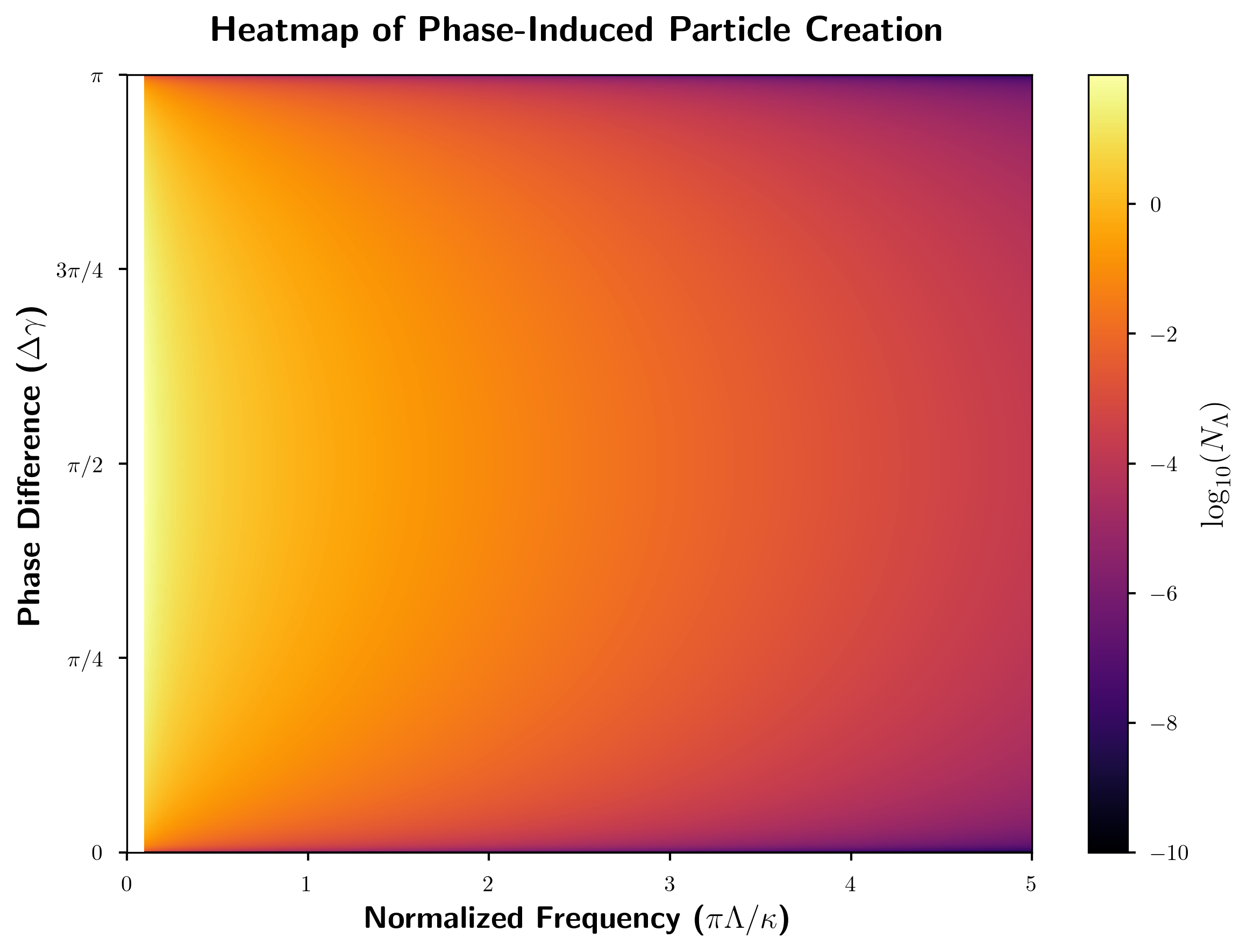}
\caption{\textbf{Heatmap of the phase-induced particle spectrum.} 
The number of created quanta $N_\Lambda = |\beta_\Lambda|^2$, from Eq.~\eqref{eq:equal-k}, is plotted on a logarithmic color scale as a function of the normalized frequency $\pi\Lambda/\kappa$ and the relative phase shift $\Delta\gamma = \gamma' - \gamma$. This visualization is for the case of equal thermal parameters ($\kappa'=\kappa$), where particle creation is driven solely by the phase mismatch. Particle creation vanishes when the frames are in phase ($\Delta\gamma = 0, \pi$) and is maximized for a phase shift of $\pi/2$. The spectrum is clearly dominated by low-frequency (infrared) modes.}
\label{fig:particle_creation_heatmap}
\end{figure}

\medskip

\subsection{Mirror realization and equivalence on \texorpdfstring{$\mathscr I^+$}{I+}}
\label{sec:bogol-mirror}

For the Carlitz–Willey ray–tracing map
\begin{align}
v=p(u)=v_H-\frac{1}{\kappa}\,e^{-\kappa u},
\label{eq:CW-map-again2}
\end{align}
the in$\to$out kernel diagonalizes in the Mellin (log–frequency) basis and yields a single–mode squeeze with thermal ratio
\begin{align}
a^{\mathrm{out}}_{\Omega}
=\alpha_\Omega\,b_{\Omega}+\beta_\Omega\,b_{\Omega}^{\dagger},
\qquad
\frac{|\beta_\Omega|^2}{|\alpha_\Omega|^2}=e^{-2\pi\Omega/\kappa},
\qquad
|\beta_\Omega|^2=\frac{1}{e^{2\pi\Omega/\kappa}-1}.
\label{eq:CW-thermal-again2}
\end{align}
Appending a chiral, frequency–diagonal phase drive near $\mathscr I^+_R$ rotates the squeeze angle without changing $|\beta_\Omega|$,
\begin{align}
\beta_\Omega\ \longrightarrow\ \beta_\Omega^{(\gamma)}=e^{\,i\,2\gamma(\Omega)}\,\beta_\Omega,\qquad
\big|\partial_\Omega\gamma(\Omega)\big|\ll \kappa^{-1}\ \text{for}\ \Omega\sim\mathcal O(\kappa).
\label{eq:phase-rot2}
\end{align}
When $\gamma(\Omega)$ is approximately flat across the thermal band one may set $\gamma(\Omega)\simeq\gamma$, and the out–state on $\mathscr I^+$ has the same Mellin–diagonal $SU(1,1)$ structure as in Eqs.~\eqref{eq:kgkg-bogol}–\eqref{eq:canonicity}. Consequently, number observables (Planck spectrum, chiral stress) agree exactly with the $\kappa\gamma$ vacuum; phase–sensitive, non–stationary observables agree whenever the flat–phase condition holds. The statement is asymptotic and chiral—on $\mathscr I^{\pm}$ there is no $u{\leftrightarrow}v$ mixing; in the bulk, mirror–induced left/right correlations may persist even when their projection on null infinity coincides.

\section{Wightman function and KMS structure}
\label{sec:wightman-kms}

We assemble the two–point function of the $\kappa\gamma$ state, display a clean
split into a stationary (thermal) part and a non–stationary, phase–dependent
part, and verify that only the former satisfies the KMS condition with respect
to Minkowski time translations. We also record the short–distance (Hadamard)
structure and a stress–tensor check. The Mellin integrals used below are
summarized in Apps.~\ref{app:integral} and \ref{app:phase_integrals}.

\subsection{Right–moving correlator on \texorpdfstring{$\mathscr I^+_R$}{I+R}}
\label{sec:wightman-kms-rtw}

Starting from the mode expansion \eqref{eq:kg_mode_field} and \eqref{eq:kg_mode}, and using $[{\cal A}_{\Lambda,\kappa,\gamma},{\cal A}^\dagger_{\Lambda',\kappa,\gamma}]
=\delta(\Lambda-\Lambda')$, the right–moving Wightman function
\begin{align}
W^{\rm RTW}_{\kappa\gamma}(u,u')
=\big\langle 0_{\kappa\gamma}\big|\Phi_{\rm RTW}(u)\,\Phi_{\rm RTW}(u')
\big|0_{\kappa\gamma}\big\rangle
\end{align}
can be written at fixed $\Lambda$ as the sum of three elementary contributions:
a thermal, $\Delta u$–dependent piece, and two phase–sensitive pieces depending
on $u{+}u'$. A short algebra gives
\begin{align}
W^{\rm RTW}_{\kappa\gamma}(u,u')
&=\int_0^\infty\frac{d\Lambda}{4\pi\,\Lambda\,\sinh(\pi\Lambda/\kappa)}\,
\Big\{
\cosh\!\Big(\frac{\pi\Lambda}{\kappa}-i\Lambda \Delta u\Big)
-\cos(2\gamma)\,\big[1-\cos\!\big(\Lambda(u{+}u')\big)\big]\nonumber\\
&\hspace{4.45cm}
+\;\sin(2\gamma)\,\sin\!\big(\Lambda(u{+}u')\big)\Big\},
\qquad \Delta u:=u-u'.
\label{eq:W-RTW-integral}
\end{align}
The “$1$” inside the square bracket carries a state–independent UV contact
term and will be removed by the standard subtraction used in App.~\ref{app:phase_integrals}.
Performing the $\Lambda$–integrals with the prescriptions in
Apps.~\ref{app:integral} and \ref{app:phase_integrals} then yields the split
\begin{empheq}[box=\widefbox]{equation}
W^{\rm RTW}_{\kappa\gamma}(u,u')
=
W^{\rm RTW}_{\rm th}(\Delta u)
\;+\;
W^{\rm RTW}_{\rm ph}(u{+}u';\gamma),
\qquad
\Delta u\equiv u-u'.
\label{eq:W-rtw-split}
\end{empheq}
The stationary (thermal) term is
\begin{equation}
W^{\rm RTW}_{\rm th}(\Delta u)
=
-\frac{1}{4\pi}\,
\ln\!\Big[\mu^2\,\frac{2}{\kappa}\,
\sinh\!\Big(\frac{\kappa}{2}\,(\Delta u-i\epsilon)\Big)\Big],
\label{eq:Wthermal-RTW}
\end{equation}
and the non–stationary phase term is
\begin{align}
W^{\rm RTW}_{\rm ph}(u{+}u';\gamma)
&=
-\frac{\cos(2\gamma)}{4\pi}\,
\ln\!\cosh\!\Big(\frac{\kappa}{2}(u{+}u')\Big)
\;+\;
\sin(2\gamma)\,I_{\sin}^{\rm (ren)}\!\big(u{+}u';\mu_{\rm IR}\big),
\label{eq:Wphase-RTW}
\end{align}
where the odd, IR–renormalized function (defined up to an odd linear scheme
choice) is
\begin{equation}
I_{\sin}^{\rm (ren)}(\Sigma;\mu_{\rm IR})
=
\frac{1}{2\pi}\sum_{n=0}^{\infty}
\!\left[
\arctan\!\frac{\kappa\,\Sigma}{\pi(2n{+}1)}
-
\frac{\kappa\,\Sigma}{\pi(2n{+}1)}
\right]
+\frac{\kappa\,\Sigma}{4\pi^2}\,
\ln\!\frac{\kappa}{2\pi\,\mu_{\rm IR}},
\qquad \Sigma\equiv u{+}u'.
\label{eq:Isin-ren}
\end{equation}
Here $i\epsilon$ fixes the Wightman boundary value; $\mu$ and $\mu_{\rm IR}$
are inert renormalization scales for the even/odd channels, respectively. The
thermal term depends only on $\Delta u$ and is thus stationary; the phase term
depends on $u{+}u'$ and is intrinsically non–stationary.

It is useful to record the short–distance expansions (Hadamard structure).
For $|\Delta u|\ll \kappa^{-1}$,
\begin{align}
W^{\rm RTW}_{\rm th}(\Delta u)
&=-\frac{1}{4\pi}\Big[\ln\!\big(\mu^2(\Delta u-i\epsilon)\big)
+\frac{\kappa^2}{24}\,(\Delta u)^2+\mathcal{O}(\Delta u)^4\Big],\\
-\frac{\cos(2\gamma)}{4\pi}\ln\!\cosh\!\Big(\frac{\kappa}{2}(u{+}u')\Big)
&=-\frac{\cos(2\gamma)}{4\pi}\Big[\frac{\kappa^2}{8}(u{+}u')^{2}
+\mathcal{O}\big((u{+}u')^{4}\big)\Big],\\
I_{\sin}^{\rm (ren)}(\Sigma;\mu_{\rm IR})
&=\frac{\kappa\,\Sigma}{4\pi^2}\ln\!\frac{\kappa}{2\pi\,\mu_{\rm IR}}
-\frac{7\zeta(3)}{48\pi^4}\,\kappa^3\Sigma^3+\mathcal{O}(\Sigma^5),
\end{align}
which shows the standard logarithmic short–distance singularity (even channel)
and the finiteness of the odd channel once the universal IR logarithm is
subtracted.

\begin{figure}[h!]
\centering
\includegraphics[width=\textwidth]{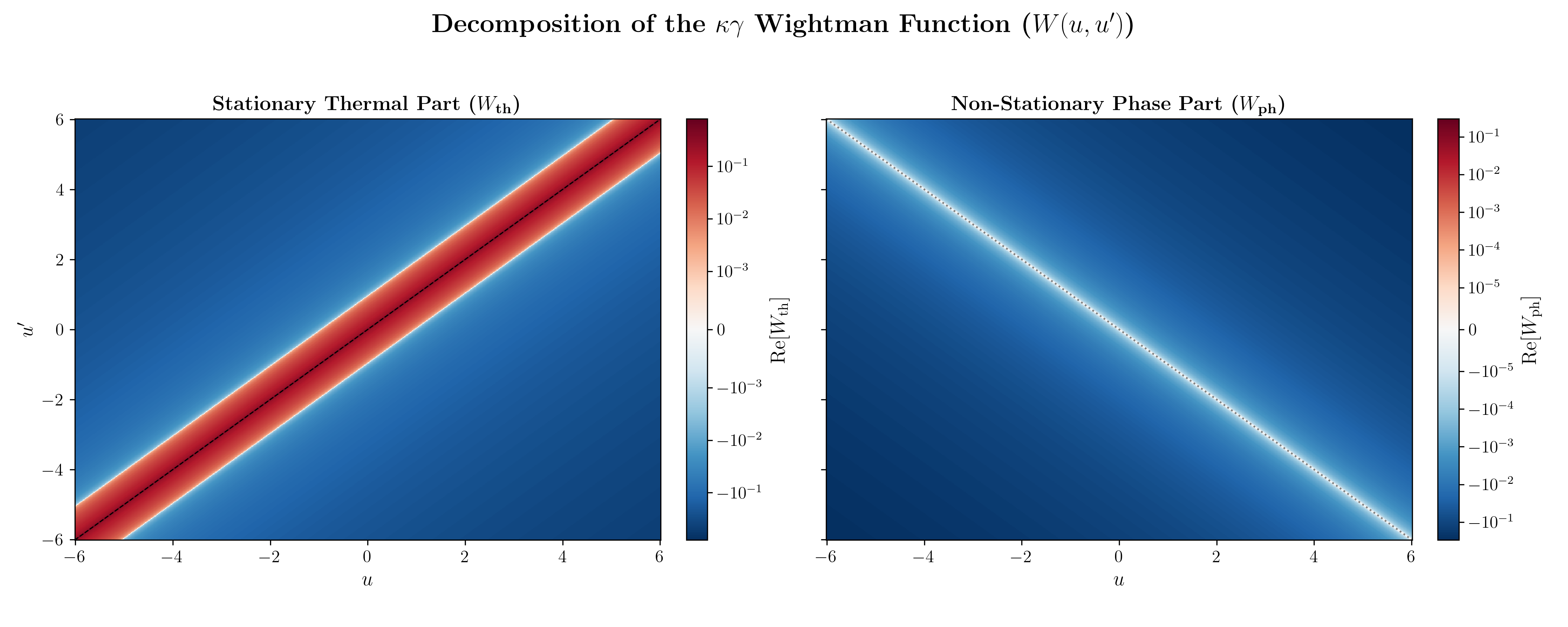}
\caption{\textbf{Decomposition of the $\kappa\gamma$ Wightman function.} 
The real part of the right-moving Wightman function $W^{\rm RTW}_{\kappa\gamma}(u, u')$ is split into its constituent parts as described in Eq.~\eqref{eq:W-rtw-split}, here plotted for $\kappa=1.0$ and $\gamma=\pi/8$. Note the separate logarithmic color scales for each panel, reflecting the different magnitudes of the two components. 
(\textbf{Left}) The stationary thermal part, $W_{\rm th}$. Its features are constant along lines of constant separation, $u-u'=\text{const}$, demonstrating its stationarity. 
(\textbf{Right}) The non-stationary phase part, $W_{\rm ph}$. Its features are constant along lines of constant average time, $u+u'=\text{const}$, demonstrating its non-stationary nature. This component is responsible for phase-sensitive phenomena.}
\label{fig:wightman_decomposition}
\end{figure}

\subsection{Left–moving sector and total correlator}
\label{sec:wightman-kms-total}

By $u\leftrightarrow v$ symmetry the left–moving correlator is obtained from
\eqref{eq:Wthermal-RTW}–\eqref{eq:Isin-ren} upon replacing $(u,u')\to(v,v')$.
Summing both chiralities gives
\begin{align}
W^{\rm tot}_{\kappa\gamma}(x,x')
&=
-\frac{1}{4\pi}\ln\!\Big[\mu^2\,\frac{2}{\kappa}\sinh\!\Big(\frac{\kappa}{2}(u-u'-i\epsilon)\Big)\Big]
-\frac{1}{4\pi}\ln\!\Big[\mu^2\,\frac{2}{\kappa}\sinh\!\Big(\frac{\kappa}{2}(v-v'-i\epsilon)\Big)\Big]\nonumber\\
&\quad
-\frac{\cos(2\gamma)}{4\pi}\!\left[
\ln\!\cosh\!\Big(\frac{\kappa}{2}(u{+}u')\Big)
+\ln\!\cosh\!\Big(\frac{\kappa}{2}(v{+}v')\Big)
\right]\nonumber\\
&\quad
+\sin(2\gamma)\!\left[
I_{\sin}^{\rm (ren)}(u{+}u';\mu_{\rm IR})
+I_{\sin}^{\rm (ren)}(v{+}v';\mu_{\rm IR})
\right].
\label{eq:Wtotal-kappagamma}
\end{align}
Two immediate limits are instructive:
(i) $\kappa\to0$ gives the standard $(1{+}1)$D Minkowski Wightman since both
phase pieces vanish ($\kappa\ln\kappa\to0$);
(ii) a uniform average over $\gamma$ (or a special choice with
$\cos(2\gamma)=\sin(2\gamma)=0$) projects out the non–stationary sector.

\subsection{KMS of the stationary sector and non–KMS of the phase sector}
\label{sec:wightman-kms-kms}

Consider an inertial worldline at fixed $x=x'$, so that $\Delta u=\Delta v=\tau$
and $S:=\tfrac{1}{2}(t+t')$. Let $W^{\gtrless}(\tau)$ denote the greater/lesser
boundary values. The KMS condition at inverse temperature $\beta_\kappa$ reads
\begin{empheq}[box=\widefbox]{equation}
W^{>}\!(\tau-i\beta_\kappa)=W^{<}\!(\tau),
\qquad \beta_\kappa=\frac{2\pi}{\kappa}.
\label{eq:KMS}
\end{empheq}
Using \eqref{eq:Wthermal-RTW} and $\sinh(z-i\pi)=-\sinh z$ shows immediately
that $W_{\rm th}$ satisfies \eqref{eq:KMS} with $\beta_\kappa=2\pi/\kappa$; hence
the stationary sector is thermal at $T_\kappa=\kappa/(2\pi)$.

The phase sector depends on $u{+}u'=2S$ (and $v{+}v'=2S$ on the other chirality),
so under $\tau\to\tau-i\beta_\kappa$ with $t'$ fixed one has
$S\to S-\tfrac{i}{2}\beta_\kappa$ and
\begin{align}
\ln\!\cosh\!\Big(\tfrac{\kappa}{2}\big[(u{+}u')-i\beta_\kappa\big]\Big)
&=\ln\!\cosh(\kappa S)+i\pi, \label{eq:cosh-shift}\\
I_{\sin}^{\rm (ren)}\!\big((u{+}u')-i\beta_\kappa\big)
&=I_{\sin}^{\rm (ren)}(2S)
-\frac{i}{2\pi}\ln\!\frac{\kappa}{2\pi\mu_{\rm IR}}+\cdots, \label{eq:Isin-shift}
\end{align}
where the ellipsis denotes nonperiodic odd functions of $S$
(App.~\ref{app:phase_integrals}). Consequently
$W_{\rm ph}^{>}(\tau-i\beta_\kappa)\neq W_{\rm ph}^{<}(\tau)$ in general: the
phase sector is intrinsically non–KMS in Minkowski time. It is invisible to
adiabatic inertial thermometry (which samples only stationary data) but can be
accessed by non–stationary probes that mix $u$ and $v$ along their worldlines
(e.g., \ accelerated UDW detectors in Sec.~\ref{sec:udw_response}).

It is sometimes convenient to restate KMS in the frequency domain for the
stationary sector. Define the Fourier transforms
\begin{align}
\widetilde W^{\gtrless}_{\rm th}(\omega)
:=\int_{-\infty}^{\infty} d\tau\,e^{i\omega\tau}\,W^{\gtrless}_{\rm th}(\tau),
\qquad
n_B(\omega):=\frac{1}{e^{\beta_\kappa\omega}-1}.
\end{align}
Analyticity of $W_{\rm th}$ in the strip $0<\Im\tau<\beta_\kappa$ implies the
detailed balance relation
\begin{empheq}[box=\widefbox]{equation}
\widetilde W^{>}_{\rm th}(\omega)=e^{\beta_\kappa\omega}\,\widetilde W^{<}_{\rm th}(\omega),
\qquad
\Rightarrow\quad
\widetilde W^{>}_{\rm th}(\omega)\ \propto\ \theta(\omega)\,[1+n_B(\omega)]+\theta(-\omega)\,n_B(-\omega),
\label{eq:KMS-spectral}
\end{empheq}
consistent with the inertial UDW response in Sec.~\ref{sec:udw_response}.

\subsection{Stress–tensor check (chiral Stefan–Boltzmann law)}
\label{sec:wightman-kms-stress}

For a massless scalar in $(1{+}1)$D, the chiral stress tensor is obtained from
the Wightman function by point splitting and subtraction of the Minkowski
vacuum,
\begin{align}
\langle T_{uu}\rangle_{\kappa\gamma}
=-\lim_{u'\to u}\Big[\partial_u\partial_{u'}W^{\rm RTW}_{\kappa\gamma}(u,u')
-\partial_u\partial_{u'}W^{\rm RTW}_{\rm Mink}(u,u')\Big].
\end{align}
Since $W^{\rm RTW}_{\rm ph}$ is independent of $\Delta u$, it does not contribute
to $\partial_u\partial_{u'}$ at coincidence. Using \eqref{eq:Wthermal-RTW} one finds
\begin{empheq}[box=\widefbox]{equation}
\langle T_{uu}\rangle_{\kappa\gamma}=\frac{\pi}{12}\,T_\kappa^2
=\frac{\kappa^2}{48\pi},
\qquad
T_\kappa=\frac{\kappa}{2\pi},
\end{empheq}
the standard chiral Stefan–Boltzmann law. The left–moving component gives the
same value, so the total energy density is $\langle T_{tt}\rangle
=\langle T_{uu}\rangle+\langle T_{vv}\rangle=\frac{\pi}{6}\,T_\kappa^2$,
as expected for a $(1{+}1)$D CFT with $c=1$.

\section{Unruh-DeWitt Detector Response}
\label{sec:udw_response}
We model the detector as a two-level Unruh-DeWitt (UDW) system \cite{Unruh1976, Einstein100} with a ground state $\ket{g}$ and an excited state $\ket{e}$, separated by an energy gap $\omega$. Its interaction with a massless scalar field $\Phi$ is described by a derivative-coupling Hamiltonian, following \cite{Svidzinsky2021PRR}:
\begin{equation}
    H_{\text{int}}(\tau) = g \frac{\partial}{\partial \tau} \Phi(\tau) \left( \sigma^\dagger e^{i \omega \tau} + \sigma e^{-i \omega \tau} \right).
    \label{eq:Hint}
\end{equation}
In this expression, $g$ represents the coupling strength, while $\sigma^\dagger = \ket{e}\bra{g}$ and $\sigma = \ket{g}\bra{e}$ are the detector's raising and lowering operators.

To find the probability of the detector becoming excited, we calculate the final state $|\Psi_f\rangle$ using first-order perturbation theory. The state after the interaction is given by:
\begin{equation}
    |\Psi_f^{(1)}\rangle = -\frac{i}{\hbar} \int_{-\infty}^{+\infty} d\tau H_{\text{int}}(\tau) |\Psi_i\rangle.
\end{equation}
Since we're interested in the detector "clicking," we focus on the component of the interaction Hamiltonian proportional to the raising operator $\sigma^\dagger$, as this term drives the transition from the ground state to the excited state. The long-time (adiabatic) transition rate can be found from the first–order excitation probability, which is given in terms of the field's Wightman function $W\big(x(\tau),x(\tau')\big)$.

We begin by defining the $\kappa\gamma$-vacuum and calculating the response of a stationary UDW detector before moving to the accelerated case. We consider a (1+1)-dimensional massless scalar field $\Phi$ for simplicity.

\subsection{Inertial Detector Response}

Let's consider an inertial UDW detector with energy gap $\omega$, held at a fixed position $x_0$. Its interaction with the field is described by first-order perturbation theory. The final state of the system, initially in $|g\rangle \otimes |0\rangle_{\kappa\gamma}$, is given by
\begin{equation}
|\Psi_f^{(1)}\rangle = -\frac{i g}{\hbar} \int_{-\infty}^{+\infty} dt \, e^{i\omega t} \sigma^\dagger  \frac{\partial}{\partial t} \Phi_{\text{RTW}}(t-x_0) |g\rangle \otimes |0\rangle_{\kappa\gamma},
\end{equation}
where we have considered only the excitation term. Projecting onto a final state with a single created particle of frequency $\Lambda$, $|1_\Lambda\rangle_{\kappa\gamma} = {\cal A}_{\Lambda,\kappa\gamma}^{\dagger} |0\rangle_{\kappa\gamma}$, we get:
\begin{align}
|\Psi_{f, \text{RTW}}^{(1)}\rangle = -\frac{i g}{\hbar} e^{i\omega x_0} \int_0^\infty d\Lambda \int_{-\infty}^{+\infty} du \, e^{i\omega u}
\frac{\partial}{\partial u}
\Phi^*_{\Lambda,\kappa,\gamma}(u) |1_\Lambda\rangle_{\kappa\gamma} |e\rangle.
\end{align}
Substituting the mode function from Eq.~\eqref{eq:kg_mode_field} and performing the integral over the light-cone coordinate $u$:
\begin{align}
\int_{-\infty}^{+\infty} du \, e^{i\omega u}
\frac{\partial}{\partial u}
\Phi^*_{\Lambda,\kappa,\gamma}(u)=& \frac{i \Lambda}{\sqrt{{\cal N}_{\Lambda,\kappa}}} \int_{-\infty}^{+\infty} du \left( e^{\frac{\pi\Lambda}{2\kappa}-i\gamma} e^{i(\omega+\Lambda)u} 
- e^{-\frac{\pi\Lambda}{2\kappa}+i\gamma} e^{i(\omega-\Lambda)u} \right) \nn\\
=& \frac{2\pi i \Lambda}{\sqrt{{\cal N}_{\Lambda,\kappa}}} \left( e^{\frac{\pi\Lambda}{2\kappa}-i\gamma} \delta(\omega+\Lambda)
-e^{-\frac{\pi\Lambda}{2\kappa}+i\gamma} \delta(\omega-\Lambda) \right),
\end{align}
Since both detector energy gap $\omega$ and mode frequency $\Lambda$ are positive, the $\delta(\omega+\Lambda)$ term vanishes. The delta function enforces energy conservation, $\Lambda = \omega$. The final state amplitude for the creation of a mode $\omega$ is then:
\begin{equation}
\langle 1_\omega, e | \Psi_{f, \text{RTW}}^{(1)} \rangle = 
- \frac{ g}{\hbar} e^{i(\omega x_0+\gamma)} {\sqrt{\pi \omega}}
\frac{e^{-\frac{\pi\omega}{2\kappa}}}
{\sqrt{2\sinh\left(\frac{\pi\omega}{\kappa}\right)}}.
\end{equation}
The transition probability per unit time, or rate $\mathcal{R}$, is proportional to the modulus squared of this amplitude.
\begin{align}
\mathcal{R}_{g \to e} \propto& \left|- \frac{ g}{\hbar} e^{i(\omega x_0+\gamma)} {\sqrt{\pi \omega}}
\frac{e^{-\frac{\pi\omega}{2\kappa}}}
{\sqrt{2\sinh\left(\frac{\pi\omega}{\kappa}\right)}}
\right|^2 \nn\\
=&  \frac{\pi \omega g^2}{\hbar^2}
\frac{1}{e^{\frac{2\pi\omega}{\kappa}}-1}.
\end{align}
Summing the contributions from right-traveling (RTW) and left-traveling (LTW) waves gives a perfect Planckian distribution. The adiabatic rate for $\omega>0$ is:
\begin{empheq}[box=\widefbox]{equation}
\dot{\mathcal F}_{\rm inertial}(\omega)\ \propto\ \frac{\omega}{2\pi}\,
\frac{1}{e^{\beta_\kappa\omega}-1},
\qquad
\beta_\kappa=\frac{2\pi}{\kappa},
\qquad (\omega>0).
\label{eq:UDW-inertial-Planck}
\end{empheq}
An inertial detector perceives the $\kappa\gamma$-vacuum as a thermal bath with temperature $T_\kappa = \kappa/(2\pi)$. The phase parameter $\gamma$ appears only as a phase in the final state amplitude and vanishes from the transition rate, meaning it is completely invisible to a stationary observer and affects only non-stationary observables. In the limit $\kappa \to 0$, the temperature goes to zero, and the detector stops clicking, as expected for the standard Minkowski vacuum.

\subsection{Uniform Acceleration: Probing the Phase Structure}

We now turn to the case of a uniformly accelerated UDW detector with proper acceleration $a>0$. This will reveal the role of the $\gamma$ parameter. An observer with constant proper acceleration follows a hyperbolic trajectory. In the right Rindler wedge, one has
\begin{align}
u(\tau)=-\frac{1}{a}e^{-a\tau},\qquad v(\tau)=\frac{1}{a}e^{+a\tau}, \qquad e^{i\omega\tau}=(-a\,u)^{-i\Omega},\qquad \Omega\equiv\frac{\omega}{a},
\end{align}
where the principal branch is used for the complex power. The detector's excitation from the vacuum involves creating a quantum with mode function $\Phi^*_{\Lambda,\kappa,\gamma}(u)$. The first-order amplitude can be written as:
\begin{align}
\left| \Psi^{(1)}_{f,\,\mathrm{RTW}} \right\rangle
=& -\frac{ig}{\hbar} \int_{-\infty}^{+\infty} d\tau \int_{0}^{\infty} d\Lambda\, e^{i\omega\tau}\,\partial_{\tau} \Phi^*\big(u(\tau), \Lambda, \kappa, \gamma\big)\,\mathcal{A}^\dagger_{\Lambda, \kappa, \gamma} \ket{0}_{\kappa\gamma}\ket{e}\nn\\
=& -\frac{ig}{\hbar} \int_{-\infty}^0 du \int_0^{\infty} d\Lambda\, \frac{1}{\sqrt{{\cal N}_{\Lambda,\kappa}}}\, (-a u)^{-i\Omega}\,
\partial_{u} \Big[ e^{\frac{\pi\Lambda}{2\kappa} - i\gamma} e^{i\Lambda u} + e^{-\frac{\pi\Lambda}{2\kappa} + i\gamma} e^{-i\Lambda u} \Big]\,
\mathcal{A}^\dagger_{\Lambda, \kappa, \gamma} \ket{0}_{\kappa\gamma}\ket{e},
\end{align}
since $d\tau\,\partial_\tau\Phi(u(\tau))=du\,\partial_u\Phi(u)$. This becomes:
\begin{align}
\left| \Psi_{f,\mathrm{RTW}}^{(1)} \right\rangle
=& \frac{g}{\hbar} \int_0^{\infty} d\Lambda\,
\frac{\Lambda\,a^{-i\Omega}}{\sqrt{{\cal N}_{\Lambda,\kappa}}}\,
\mathcal{A}^\dagger_{\Lambda, \kappa, \gamma} \ket{0}_{\kappa\gamma}\ket{e}\nn\\
&\times\Bigg\{ e^{\frac{\pi\Lambda}{2\kappa} - i\gamma} \int_{-\infty}^0 du\, (-u)^{-i\Omega} e^{i\Lambda u}
- e^{-\frac{\pi\Lambda}{2\kappa} + i\gamma} \int_{-\infty}^0 du\, (-u)^{-i\Omega} e^{-i\Lambda u} \Bigg\}.
\end{align}
Using the standard integrals (for $\Lambda>0$, principal branches)
\begin{align}
\int_{-\infty}^0 du \, (-u)^{-i\Omega} e^{+i\Lambda u}&=-i\,\Lambda^{-1+i\Omega}\,e^{-\frac{\pi\Omega}{2}}\,\Gamma(1-i\Omega),\nn\\
\int_{-\infty}^0 du \, (-u)^{-i\Omega} e^{-i\Lambda u}&=+i\,\Lambda^{-1+i\Omega}\,e^{+\frac{\pi\Omega}{2}}\,\Gamma(1-i\Omega),
\end{align}
we find the final state can be written as
\begin{align}
\left| \Psi_{f,\mathrm{RTW}}^{(1)} \right\rangle
=& -\frac{ig}{\hbar} \int_0^{\infty} d\Lambda\,
\frac{\Lambda\,a^{-i\Omega}}{\sqrt{{\cal N}_{\Lambda,\kappa}}}\,\ket{1_{\Lambda}}_{\kappa\gamma}\ket{e} \times \Big(
e^{\frac{\pi\Lambda}{2\kappa} - i\gamma}e^{-\frac{\pi\Omega}{2}}
+ e^{-\frac{\pi\Lambda}{2\kappa} + i\gamma}e^{+\frac{\pi\Omega}{2}}
\Big)\,\Lambda^{-(1-i\Omega)}\,\Gamma(1 - i\Omega) \nn \\
=& -\frac{2ig}{\hbar} \int_0^{\infty} d\Lambda\,
\frac{(\Lambda/a)^{i\Omega}\,\Gamma(1-i\Omega)}{\sqrt{{\cal N}_{\Lambda,\kappa}}}\,
\cosh\!\left[\frac{\pi}{2}\!\left(\frac{\Lambda}{\kappa}-\frac{\omega}{a}\right)-i\gamma\right]
\ket{1_{\Lambda}}_{\kappa\gamma}\ket{e}.
\end{align}
The transition rate for creating a mode of frequency $\Lambda$, denoted ${\cal R}_{g \to e}(\Lambda)$, is proportional to the squared modulus of the integrand. Using $|(\Lambda/a)^{i\Omega}|=1$, $|\Gamma(1-ix)|^2=\pi x/\sinh(\pi x)$, and $|\cosh(A-iB)|^2=\big(\cosh(2A)+\cos(2B)\big)/2$, we obtain the per-mode RTW rate:
\begin{empheq}[box=\widefbox]{equation}
{\cal R}^{\rm RTW}_{g \to e}(\Lambda)
=\frac{g^2\omega}{2\hbar^2 a\,\Lambda}\;
\frac{\cosh\!\Big(\pi\big(\frac{\Lambda}{\kappa}-\frac{\omega}{a}\big)\Big)+\cos(2\gamma)}
{2\,\sinh\!\big(\frac{\pi\Lambda}{\kappa}\big)\,\sinh\!\big(\frac{\pi\omega}{a}\big)}.
\label{eq:UDW-accel-rate}
\end{empheq}
The left–moving contribution is analogous; the total is their sum. Unlike the inertial case, this rate is not a simple Planckian spectrum and depends on both vacuum parameters $\kappa$ and $\gamma$. The phase $\gamma$, hidden from the inertial observer, now appears as an interference term $\cos(2\gamma)$, modulating the response. The accelerated motion of the detector allows it to probe the relative phase structure of the vacuum state.

\subsubsection{Destructive Interference and Detector Silence}
A remarkable feature of the transition rate in Eq.~\eqref{eq:UDW-accel-rate} is the possibility for it to vanish completely for a specific mode $\Lambda$, effectively silencing the detector's response. This occurs if the numerator becomes zero, which requires two conditions to be met simultaneously: \textbf{Frequency Matching}, where $\Lambda/\kappa = \omega/a$, and \textbf{Phase Opposition}, where $\cos(2\gamma) = -1$.

\begin{figure}[h!]
    \centering
    \includegraphics[width=\textwidth]{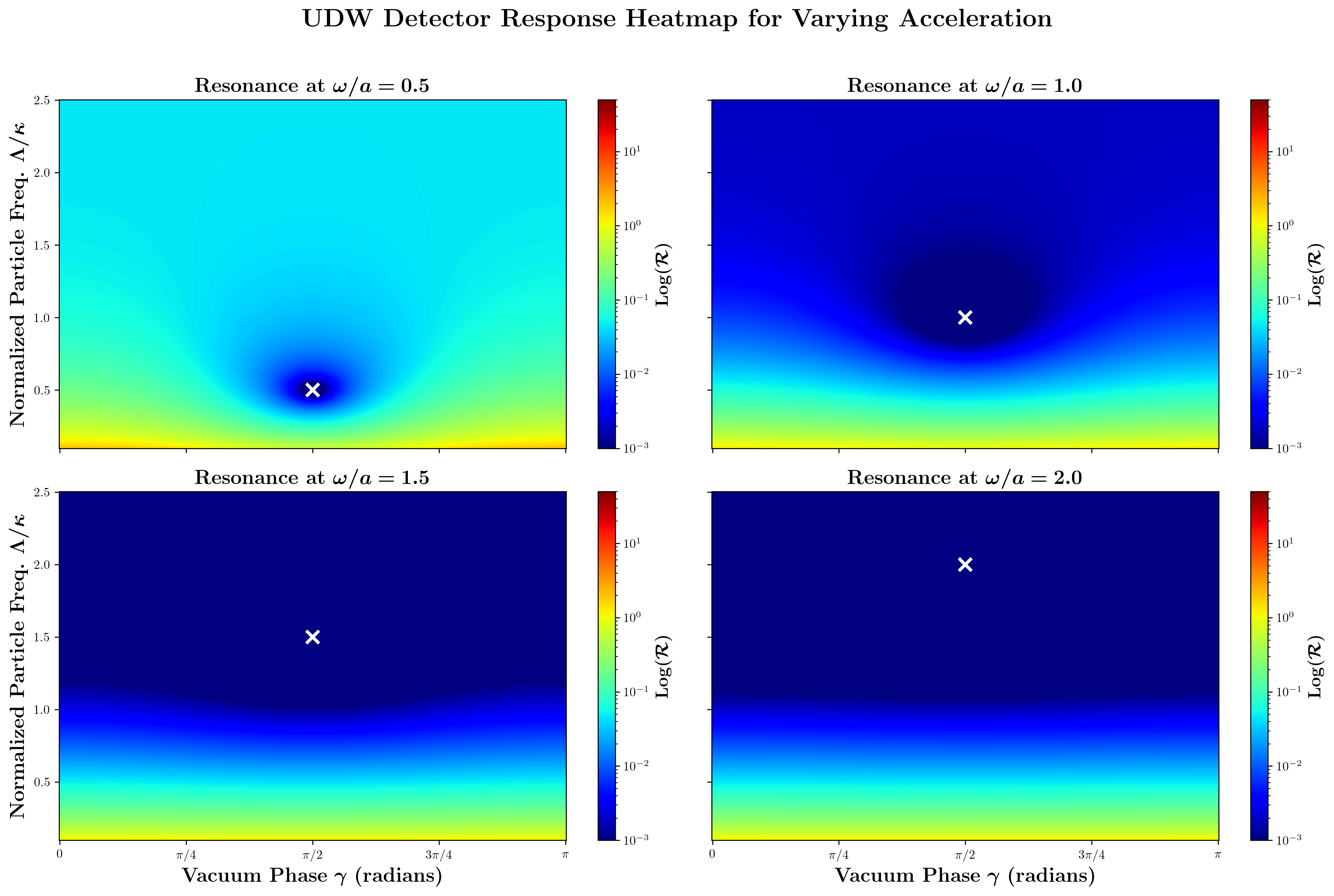}
    \caption{\textbf{Heatmap visualization of the UDW detector response.} The normalized transition rate, $\mathcal{R}(\Lambda, \gamma)$, is plotted on a logarithmic color scale as a function of the vacuum phase $\gamma$ (horizontal axis) and the normalized particle frequency $\Lambda/\kappa$ (vertical axis). Each of the four panels corresponds to a different detector resonance, set by the ratio of the detector's energy gap to its acceleration, $\omega/a$. The central feature is the point of complete response suppression (marked by a white 'x'), where destructive quantum interference drives the transition rate to zero. As predicted by the condition $\Lambda/\kappa = \omega/a$, the location of this "detector silence" is tunable, shifting vertically as the detector's acceleration changes.}
    \label{fig:udw_heatmap_panels}
\end{figure}

When both conditions are fulfilled, the detector's excitation probability for that mode is zero. This phenomenon, a direct manifestation of destructive quantum interference, is visualized in the heatmaps of Fig.~\ref{fig:udw_heatmap_panels}. The plots confirm that the "detector silence" (marked by a white 'x') occurs precisely at $\gamma=\pi/2$ and that its spectral location, $\Lambda/\kappa$, is tunable via the detector's acceleration through the ratio $\omega/a$. This highlights the physical nature of $\gamma$ as a control parameter that can be used to suppress quantum interactions.

This fine-tuned cancellation has a clear physical meaning. A uniformly accelerated Unruh–DeWitt detector has a vanishing excitation rate in the Rindler vacuum; here, the same silence is reproduced \emph{for a specific spectral line} by phase-tuning the $\kappa\gamma$ state. The analogy has limits: in the Rindler vacuum the null response holds across all energy gaps, whereas here the cancellation is mode-selective. Consequently, the \emph{integrated} rate is generically nonzero unless the detector’s spectral response is sharply localized around the canceling mode.

The cancellation is also structurally stable. As shown in Fig.~\ref{fig:udw_heatmap_panels}, any detuning from the precise conditions restores a small but positive rate. Writing $\Delta\equiv \Lambda/\kappa-\omega/a$ and $\gamma=\frac{\pi}{2}+\delta\gamma$, one finds that for small deviations, the numerator grows quadratically from zero:
\[
\cosh\!\big(\pi\Delta\big)+\cos(2\gamma)
\;\approx\;
\frac{\pi^2}{2}\,\Delta^2+2\,(\delta\gamma)^2\ \ge 0.
\]
Non-adiabatic switching would further broaden the detector’s spectral window and smear the exact zero unless the interaction time is long enough to resolve the canceling mode.

\subsubsection{The Minkowski Limit and the Unruh Effect}
We now verify that we recover the standard Unruh effect in the Minkowski vacuum limit, which corresponds to $\kappa \to 0$. For a fixed $\Lambda>0$:
\begin{align}
\sinh\!\left(\frac{\pi\Lambda}{\kappa}\right)\to\frac{1}{2}e^{\frac{\pi\Lambda}{\kappa}},\qquad
\cosh\!\left[\pi\!\left(\frac{\Lambda}{\kappa}-\frac{\omega}{a}\right)\right]\sim \frac{1}{2}e^{\frac{\pi\Lambda}{\kappa}}e^{-\frac{\pi\omega}{a}}.
\end{align}
The bounded $\cos(2\gamma)$ term is negligible against the exponentially growing terms. Hence:
\begin{align}
\lim_{\kappa\to 0} {\cal R}^{\rm RTW}_{g \to e}(\Lambda)
= \frac{g^2\omega}{2\hbar^2 a\,\Lambda}\,\frac{1}{e^{\frac{2\pi\omega}{a}}-1}.
\end{align}
This is the Unruh result: a Bose–Einstein factor at the Unruh temperature $T_U=a/(2\pi)$.

\subsubsection{Remarks on Switching and Integration}
The expressions above are spectral in the created particle frequency $\Lambda$. A physical response with smooth switching would weight the rate in Eq.~\eqref{eq:UDW-accel-rate} and its LTW counterpart and, depending on the observable, may integrate over $\Lambda$ with a detector bandwidth. The key takeaway is that inertial adiabatic responses project out the non-stationary sector of the field and are exactly thermal at $T_\kappa$, while uniformly accelerated responses retain sensitivity to the phase $\gamma$ and the ratio $\kappa/a$.

\subsection{\texorpdfstring{CW mirror with Robin boundary at $\mathscr I^+_R$}{}}
\label{sec:udw_CW_Robin}

At future null infinity, the Carlitz--Willey (CW) trajectory produces a chiral, stationary thermal kernel for right movers. A time–dependent Robin interaction on the mirror worldline modifies only the \emph{phase} of the outgoing single–mode squeeze while preserving the Planckian modulus fixed by $\kappa$. On $\mathscr I^+_R$ this is captured by the same right–moving Wightman decomposition used in Sec.~\ref{sec:wightman-kms-rtw},
\begin{empheq}[box=\widefbox]{equation}
W^{\rm RTW}_{\rm CW}(u,u')
=
W^{\rm RTW}_{\rm th}(\Delta u)
\;+\;
W^{\rm RTW}_{\rm ph}(u{+}u';\gamma_{\mathrm{eff}}),
\qquad
\Delta u:=u-u',
\label{eq:W-CW-Robin}
\end{empheq}
with $W_{\rm th}$ given by \eqref{eq:Wthermal-RTW} and $W_{\rm ph}$ by \eqref{eq:Wphase-RTW}, but with the squeeze angle replaced by an \emph{effective} phase $\gamma_{\mathrm{eff}}$. Physically, $\gamma_{\mathrm{eff}}$ is fixed by the phase of the Robin drive (Appendix~\ref{app:robin_boundary}) once projected onto the Mellin band populated by the CW flux. When the drive is weak and its phase is approximately flat over that band, $\gamma_{\mathrm{eff}}$ can be taken as frequency–independent; otherwise one may regard $\gamma_{\mathrm{eff}}$ as slowly varying in $\Lambda$ without altering the discussion below. The left–moving sector follows by $u\!\leftrightarrow v$ and is added at the end.

\subsubsection{\texorpdfstring{Inertial detector at $\mathscr I^+_R$}{}}

For a stationary inertial worldline at fixed $x_0$ one has $u=t-x_0$, so the adiabatic response of a derivative–coupled UDW detector depends only on the \emph{stationary} part of the kernel:
\[
\dot{\mathcal F}^{\rm CW}_{\rm inertial}(\omega)
=\int_{-\infty}^{\infty}\!d\Delta t\,e^{-i\omega\Delta t}\,
\partial_t\partial_{t'}\,W^{\rm RTW}_{\rm th}(t{-}x_0,t'{-}x_0),
\]
because the non–stationary piece $W_{\rm ph}(u{+}u')$ averages out under adiabatic sampling. Evaluating the standard integral yields the Planck law at $T_\kappa=\kappa/(2\pi)$, independent of the Robin phase:
\begin{empheq}[box=\widefbox]{equation}
\dot{\mathcal F}^{\rm CW}_{\rm inertial}(\omega)\ \propto\ \frac{\omega}{2\pi}\,
\frac{1}{e^{\beta_\kappa\omega}-1},
\qquad
\beta_\kappa=\frac{2\pi}{\kappa},
\qquad (\omega>0),
\label{eq:UDW-inertial-CW-Robin}
\end{empheq}
in agreement with \eqref{eq:UDW-inertial-Planck}. Thus the Robin modulation is invisible to stationary inertial thermometry on $\mathscr I^+_R$.

\subsubsection{\texorpdfstring{Uniformly accelerated detector at $\mathscr I^+_R$}{}}

A uniformly accelerated worldline in the right wedge,
\[
u(\tau)=-a^{-1}e^{-a\tau},\qquad v(\tau)=a^{-1}e^{a\tau},
\]
mixes $u$ along the trajectory and therefore probes the \emph{phase–sensitive} sector $W_{\rm ph}$. The adiabatic response for derivative coupling is obtained by pulling back \eqref{eq:W-CW-Robin}:
\[
\dot{\mathcal F}^{\rm CW}_{\rm acc}(\omega)
=\!\int\! d\tau\,d\tau'\,e^{-i\omega(\tau-\tau')}\,
\partial_\tau\partial_{\tau'}\,
W^{\rm RTW}_{\rm CW}\!\big(u(\tau),u(\tau')\big),
\]
and evaluating the same Mellin integrals used in Sec.~\ref{sec:udw_response}. The result is the per–mode right–moving rate
\begin{empheq}[box=\widefbox]{equation}
{\cal R}^{\rm CW}_{g \to e}(\Lambda)
=\frac{g^2\omega}{2\hbar^2 a\,\Lambda}\;
\frac{\displaystyle
\cosh\!\Big(\pi\Big(\frac{\Lambda}{\kappa}-\frac{\omega}{a}\Big)\Big)
+\cos\!\big(2\gamma_{\mathrm{eff}}(\Lambda)\big)}
{2\,\sinh\!\big(\frac{\pi\Lambda}{\kappa}\big)\,\sinh\!\big(\frac{\pi\omega}{a}\big)}.
\label{eq:UDW-accel-rate-CW-Robin-W}
\end{empheq}
(Left movers add analogously; the total is their sum.)

Two consequences are immediate. First, for the unmodulated CW mirror (or Neumann $f_0=0$) one has $\gamma_{\mathrm{eff}}=0$ so $\cos(2\gamma_{\mathrm{eff}})=1$, and the numerator in \eqref{eq:UDW-accel-rate-CW-Robin-W} is $\ge 2$, with equality only at $\Lambda/\kappa=\omega/a$; hence there is no destructive–interference zero and the accelerated response is strictly positive for $\Lambda>0$, reproducing \eqref{eq:UDW-accel-rate}. Second, with a Robin modulation that yields $\gamma_{\mathrm{eff}}(\Lambda_\star)=\pi/2\ (\mathrm{mod}\ \pi)$ at the frequency–matching point $\Lambda_\star/\kappa=\omega/a$, the numerator vanishes and the rate for that spectral line is zero. Small detunings restore a positive rate quadratically:
\[
\cosh\!\big(\pi\Delta\big)+\cos\!\big(2\gamma_{\mathrm{eff}}\big)
\approx \frac{\pi^2}{2}\,\Delta^2 + 2\,(\delta\gamma_{\mathrm{eff}})^2,
\qquad
\Delta:=\frac{\Lambda}{\kappa}-\frac{\omega}{a}.
\]

\paragraph{Remarks on frequency dependence.}
If the Robin drive induces an angle $\gamma_{\mathrm{eff}}(\Lambda)$ that varies slowly across the thermal band $\Lambda\sim\mathcal O(\kappa)$, one may take $\gamma_{\mathrm{eff}}$ as constant and \eqref{eq:UDW-accel-rate-CW-Robin-W} reduces to \eqref{eq:UDW-accel-rate} with $\gamma\!\to\!\gamma_{\mathrm{eff}}$. When the variation is appreciable, \eqref{eq:UDW-accel-rate-CW-Robin-W} predicts a spectrally structured accelerated response, while the inertial rate \eqref{eq:UDW-inertial-CW-Robin} remains exactly Planckian.

\section{Dynamical origin: a modulated Carlitz--Willey mirror with a null–infinity phase drive}
\label{sec:dynamical_origin_merged}

The purely thermal $\kappa$ plane–wave vacuum admits an operational realization in terms of an accelerating Carlitz--Willey (CW) mirror on future null infinity $\mathscr I^+$: the ray–tracing map $v=p(u)=v_H-\kappa^{-1}e^{-\kappa u}$ diagonalizes the in$\to$out kernel in a Mellin (log–frequency) basis so that each Mellin mode undergoes a single–mode squeeze with a Planck ratio fixed by $\kappa$. Concretely,
\begin{equation}
\label{eq:CW-thermal-merged}
a^{\text{out}}_{\Omega}=\alpha_{\Omega}\,b_{\Omega}+\beta_{\Omega}\,b^{\dagger}_{\Omega},
\qquad
\frac{|\beta_{\Omega}|^2}{|\alpha_{\Omega}|^2}=e^{-2\pi\Omega/\kappa},
\qquad
|\beta_{\Omega}|^2=\frac{1}{e^{2\pi\Omega/\kappa}-1}.
\end{equation}
This implements the $\gamma=0$ member of the $\kappa\gamma$ family: the temperature $T_{\kappa}=\kappa/(2\pi)$ is set kinematically by the CW trajectory, while the squeeze \emph{angle} is fixed to a reference value.

To obtain the full $\kappa\gamma$ state one must endow the mirror with additional dynamics. The key point is that the phase $\gamma$ does not arise from changing the trajectory $p(u)$ (which already fixes the thermal kernel), but from modulating the \emph{boundary interaction} in a chiral and frequency–diagonal way. A concrete bulk description is a time–dependent Robin/impedance law on the worldline, $(\partial_n+f(\tau))\Phi|_{\text{worldline}}=0$, with $f(\tau)$ carrying a controlled oscillatory component. In the asymptotic (null–infinity) picture it is more transparent to replace the reflecting boundary by a transparent, right–moving ``phase plate'' localized near $\mathscr I^+_R$ that couples quadratically to the Mellin modes. A minimal model is (see appendix~\ref{app:chiral_pump})
\begin{equation}
\label{eq:chiral-pump-merged-body}
H_{\text{int}}^{\rm RTW}(u)=\frac{i}{2}\!\int_{0}^{\infty}\!d\Omega\,
\Big[\zeta(u)\,b_{\Omega}^{\dagger\,2}-\zeta^*(u)\,b_{\Omega}^{2}\Big],
\end{equation}
with a slowly varying envelope $\zeta(u)$. At leading order this drive rotates the squeeze angle without changing the thermal modulus $|\beta_\Omega|$; over the thermally populated band $\Omega\sim\mathcal O(\kappa)$, if the phase of the pump is approximately flat in frequency,
\[
\arg\!\int du\,\zeta(u)\,e^{2i\Omega u}\approx \text{const}\equiv 2\gamma\quad
\text{for }\ \Omega\in[c_1\kappa,c_2\kappa],
\]
then the outgoing Bogoliubov coefficient acquires a constant phase,
\begin{equation}
\label{eq:beta-phase}
\beta_{\Omega}\ \longrightarrow\ \beta_{\Omega}^{(\gamma)}=e^{\,i\,2\gamma}\,\beta_{\Omega},
\end{equation}
and the single–mode map on $\mathscr I^+_R$ becomes
\begin{equation}
\label{eq:angle-rot}
a^{(\gamma)}_{\Omega}=\alpha_{\Omega}\,b_{\Omega}+e^{i\,2\gamma}\,\beta_{\Omega}\,b^{\dagger}_{\Omega}.
\end{equation}
Because the drive is chiral (right–moving here; left–moving obtained by $u\!\to\!v$), there is no $u$–$v$ mixing, and the magnitude $|\beta_\Omega|$ remains exactly Planckian as in \eqref{eq:CW-thermal-merged}.

Transforming back from Mellin to frequency variables, one recovers the $\kappa\gamma$ mode expansion with coefficients $e^{\frac{\pi\Lambda}{2\kappa}\pm i\gamma}$ and normalization ${\cal N}_{\Lambda,\kappa}=8\pi\Lambda\sinh(\pi\Lambda/\kappa)$. Consequently, on $\mathscr I^+$ the state produced by a CW trajectory followed by the chiral phase drive \eqref{eq:chiral-pump-merged-body} coincides with the $\kappa\gamma$ vacuum: all number observables are identical to the CW ($\gamma=0$) ones, while phase–sensitive observables (the non–stationary pieces of the Wightman function and accelerated UDW response) reproduce the $\gamma$ dependence derived earlier. The identification is intrinsically asymptotic (it holds at null infinity); in the bulk, the mirror’s left/right mixing persists and is only projected out in the $\mathscr I^\pm$ limit.

In summary, $\kappa$ is kinematic and set by the CW ray–tracing map, whereas $\gamma$ is operational and set by a null–localized, frequency–flat phase of a weak quadratic pump (or equivalently by the phase of a time–dependent Robin plate). This dynamical separation explains why the thermal sector is strictly KMS at $T_\kappa$ while the phase sector is non–stationary yet controlled, and it supplies a concrete route to engineer $\kappa\gamma$ vacua in moving–mirror analogs.

\begin{figure}[h!]
\centering
\includegraphics[width=0.8\textwidth]{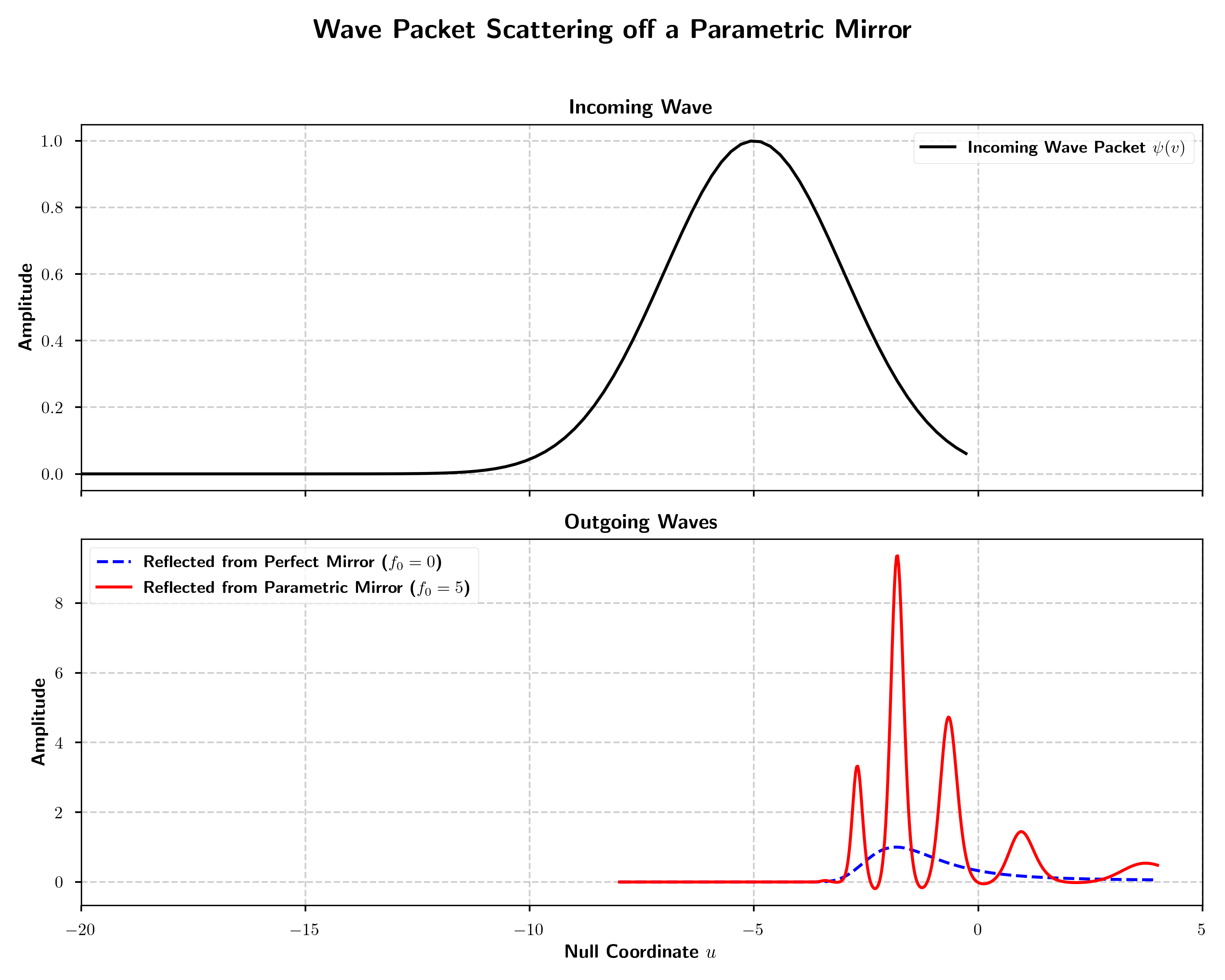}
\caption{\textbf{Numerical simulation of a Gaussian wave packet scattering off an accelerating Carlitz-Willey mirror.} 
The top panel shows the incident, right-moving Gaussian wave packet $\psi(v)$ with unit amplitude. 
The bottom panel compares the resulting left-moving outgoing wave packet $\phi(u)$ for two different boundary conditions. 
The \textcolor{blue}{blue dashed line} shows the reflection from a perfect mirror with a simple Neumann boundary condition ($f_0=0$), where the packet's shape and amplitude are largely preserved. 
The \textcolor{red}{red solid line} shows the reflection from a parametric mirror with an oscillating boundary impedance ($f_0=5.0$). 
The outgoing packet is heavily modulated and exhibits strong \textbf{parametric amplification}, with its peak amplitude significantly exceeding that of the incident wave. This demonstrates how the time-dependent boundary condition $f(\tau)$ actively pumps energy into the field, generating the complex structure associated with the $\gamma$ parameter.}
\label{fig:mirror_scattering}
\end{figure}

\begin{figure}[h!]
\centering
\includegraphics[width=\textwidth]{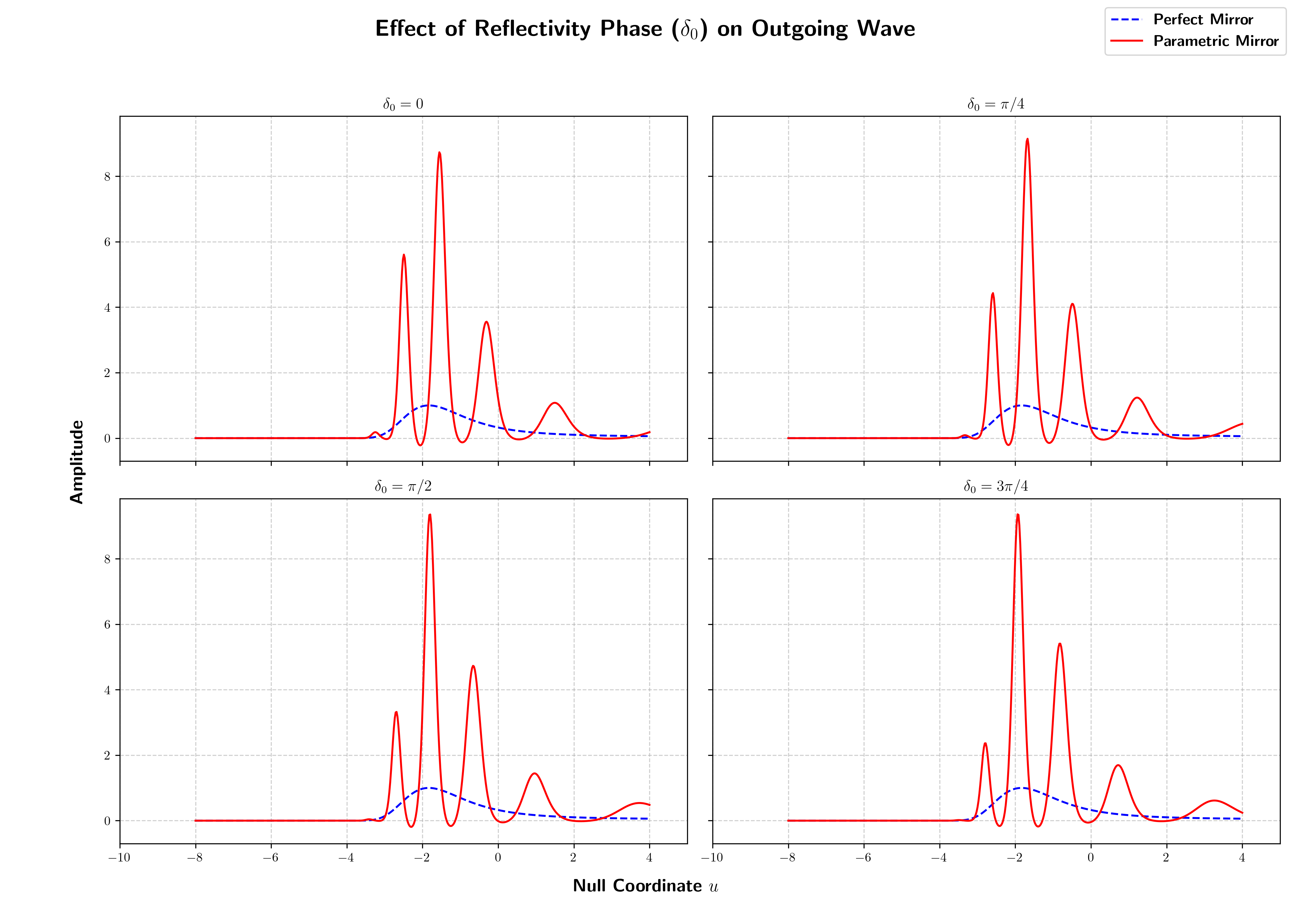}
\caption{\textbf{Effect of the reflectivity phase ($\delta_0$) on the scattered wave packet.} 
Each panel shows the outgoing wave packet $\phi(u)$ from a Gaussian incident on a Carlitz-Willey mirror. The \textcolor{blue}{blue dashed line} is the reflection from a perfect mirror ($f_0=0$), serving as a baseline. The \textcolor{red}{red solid line} is the reflection from a parametric mirror with an oscillating boundary impedance $f(\tau) \propto \cos(\omega_0\tau + \delta_0)$. Each panel corresponds to a different value of the pump phase $\delta_0$. The plot clearly demonstrates that the phase $\delta_0$ primarily controls the \textbf{position (horizontal shift)} of the reflected wave packet along the $u$-axis, while also subtly altering its internal interference pattern and peak amplitudes. This provides a dynamical visualization of how the $\gamma$ parameter shapes the structure of the outgoing quantum state.}
\label{fig:phase_comparison}
\end{figure}

\subsection{\texorpdfstring{$\kappa \to 0$ limit}{}}

For the Carlitz–Willey (CW) mirror the outgoing and ingoing null coordinates are related by the exponential ray–tracing map
\begin{equation}
  v \equiv p(u) = -\frac{1}{\kappa}\,e^{-\kappa u},
  \label{eq:CW-pu}
\end{equation}
which yields a stationary thermal flux at temperature \(T=\kappa/(2\pi)\) \cite{Carlitz_Willey1987,Birrell_Davies1982,Fulling_Davies1977,Fulling_Davies1976}. In the small–\(\kappa\) regime, a straightforward expansion shows
\begin{equation}
  p(u) = -\frac{1}{\kappa}\Big(1-\kappa u+\tfrac{1}{2}\kappa^2 u^2+\cdots\Big)
       = u - \frac{1}{\kappa} + \mathcal O(\kappa u^2),
  \qquad
  p'(u)=e^{-\kappa u}=1-\kappa u+\mathcal O(\kappa^2),
  \label{eq:affine-limit}
\end{equation}
so up to a large constant translation in \(v\) the map becomes affine, i.e., asymptotically the identity on \(\mathscr I^{+}\).

The renormalized energy flux at future null infinity is governed by the Schwarzian derivative of \(p(u)\),
\begin{equation}
  F(u) = -\frac{1}{24\pi}\,\{p(u),u\},
  \qquad
  \{p,u\} \equiv \frac{p''' }{p'} - \frac{3}{2}\Big(\frac{p''}{p'}\Big)^{\!2},
  \label{eq:flux}
\end{equation}
and for \eqref{eq:CW-pu} one finds \(p'=e^{-\kappa u}\), \(p''=-\kappa e^{-\kappa u}\), \(p'''=\kappa^2 e^{-\kappa u}\), hence
\begin{equation}
  \{p,u\} = \kappa^2 - \frac{3}{2}\kappa^2 = -\frac{\kappa^2}{2}
  \quad \Rightarrow \quad
  F(u)=\frac{\kappa^2}{48\pi}\,.
\end{equation}
Taking \(\kappa\to 0\) gives
\begin{equation}
  \lim_{\kappa\to 0}F(u)=0,
  \qquad
  \lim_{\kappa\to 0}T=\frac{\kappa}{2\pi}=0,
  \qquad
  \lim_{\kappa\to 0}\langle N_{\omega}\rangle=\frac{1}{e^{2\pi\omega/\kappa}-1}=0,
\end{equation}
so the out–state reduces to vacuum with no particle creation. This is the expected Möbius–invariance statement: the affine limit \eqref{eq:affine-limit} has vanishing Schwarzian and therefore zero flux \cite{Birrell_Davies1982,Good2013mirror}.

The worldline interpretation is consistent. With \(u=t-x\) and \(v=t+x\), the affine limit \(v\simeq u - 1/\kappa\) implies \(t+x \simeq t-x + \text{const}\), hence \(x\simeq \text{const}/2\): the mirror becomes inertial and static after a pure null translation of \(v\). Since such translations do not affect Schwarzian invariants, all physically measurable quantities on \(\mathscr I^{+}\) smoothly approach their Minkowski–vacuum values in the \(\kappa\to 0\) limit \cite{Birrell_Davies1982}.

\section{Conclusion}
\label{sec:conclusion}

This work ties a clean physical bow on the kinematic $\kappa\gamma$ family. An accelerating Carlitz–Willey mirror fixes the thermal weights, while a gentle, chiral boundary modulation—equivalently a time-dependent Robin impedance—acts as a phase plate that rotates the squeeze angle without disturbing the Planck modulus to leading order. On future null infinity the state therefore splits neatly into a stationary thermal sector and a controlled, phase-sensitive, non-stationary sector. Inertial probes see an exact Planck law at temperature set by $\kappa$; uniformly accelerated probes mix along their worldlines and expose $\gamma$ through interference, including a striking mode-selective suppression when the drive and detector are frequency matched. Numerical wave-packet simulations visualize the same mechanism, showing the phase imprint and parametric amplification induced by the modulation.

The conceptual message is simple: trajectory sets scale, boundary sets angle. This separation clarifies why thermodynamic observables are strictly thermal while phase-sensitive ones are not, and it furnishes an operational recipe to engineer the full $\kappa\gamma$ vacuum in moving-mirror analogs. Beyond establishing this realization, the framework opens several avenues: allowing a gentle frequency dependence of the angle, exploring stronger drives where angle–modulus leakage becomes relevant, studying left–right mixing away from null infinity, and translating these ideas to platforms such as superconducting circuits, optomechanics, and metamaterials where dynamic boundary control is natural.

In short, modulated accelerating mirrors provide a practical and flexible route from abstract squeeze parameters to laboratory-tunable signatures. With $\kappa$ anchored by geometry and $\gamma$ dialed by boundary pumping, the phase structure of quantum radiation becomes a controllable resource rather than a byproduct.

\section*{Acknowledgments}

I am grateful to Girish Agarwal, Marlan Scully, Bill Unruh, and Suhail Zubairy for discussions.  

This work was supported by the
Robert A. Welch Foundation (Grant No. A-1261) and the National Science Foundation (Grant No. PHY-2013771).

\appendix 

\section{Evaluation of the Thermal Wightman Integral} \label{app:integral}
Let $\Delta u=u-u'$. We evaluate
\begin{align}
I
&=\int_{0}^{\infty}\frac{d\Lambda}{4\pi\,\Lambda\,\sinh\!\left(\frac{\pi\Lambda}{\kappa}\right)}
\,\cosh\!\left(\frac{\pi\Lambda}{\kappa}-i\Lambda\Delta u\right).
\label{eq:integral_start}
\end{align}
Differentiating with respect to $\Delta u$ removes the explicit $1/\Lambda$:
\begin{align}
\partial_{\Delta u} I
&=-\frac{i}{4\pi}\int_{0}^{\infty}\!d\Lambda\,
\frac{\sinh\!\left(\frac{\pi\Lambda}{\kappa}-i\Lambda\Delta u\right)}
{\sinh\!\left(\frac{\pi\Lambda}{\kappa}\right)}
\label{eq:dI_start}\\
&=-\frac{1}{4\pi}\int_{0}^{\infty}\!d\Lambda\,\coth\!\left(\frac{\pi\Lambda}{\kappa}\right)
\sin\!\big(\Lambda(\Delta u-i\epsilon)\big)\;-\;\frac{i}{4\pi}\int_{0}^{\infty}\!d\Lambda\,\cos\!\big(\Lambda(\Delta u-i\epsilon)\big),
\label{eq:dI_coth}
\end{align}
where we used $\sinh(a-ib)=\sinh a\cos b-i\cosh a\sin b$ and introduced $i\epsilon$ for the Wightman prescription. The second integral in \eqref{eq:dI_coth} is a contact term $\propto\delta(\Delta u)$ and can be dropped for $\Delta u\neq 0$.

To obtain the final expression for the integral $I$, we begin by integrating the result for its derivative with respect to $\Delta u$, as given in Eq.~(\ref{eq:dI_coth}). For the region where $\Delta u \neq 0$, the dominant term is
\begin{equation}
\partial_{\Delta u} I = -\frac{1}{4\pi}\int_{0}^{\infty}\!d\Lambda\,\coth\!\left(\frac{\pi\Lambda}{\kappa}\right) \sin\!\big(\Lambda(\Delta u-i\epsilon)\big).
\end{equation}
The integral is a standard Fourier sine transform of the hyperbolic cotangent, which evaluates to
\begin{equation}
\int_{0}^{\infty}\!d\Lambda\,\coth\!\left(\frac{\pi\Lambda}{\kappa}\right) \sin\!\big(\Lambda b\big) = \frac{\kappa}{2} \coth\left(\frac{\kappa b}{2}\right),
\end{equation}
where $b = \Delta u - i\epsilon$. Substituting this result back yields a simplified expression for the derivative,
\begin{equation}
\partial_{\Delta u} I = -\frac{\kappa}{8\pi} \coth\left(\frac{\kappa}{2}(\Delta u - i\epsilon)\right).
\end{equation}
This expression can be directly integrated with respect to $\Delta u$. Using the identity $\int \coth(x) dx = \ln(\sinh(x))$, we find
\begin{equation}
I = -\frac{1}{4\pi} \ln\left[\sinh\left(\frac{\kappa}{2}(\Delta u - i\epsilon)\right)\right] + C,
\end{equation}
where $C$ is a constant of integration. This constant is fixed by requiring that in the limit $\kappa \to 0$, the expression correctly reproduces the right-moving $1+1$ Minkowski vacuum correlator, which is proportional to $\ln(\Delta u - i\epsilon)$. This condition sets the constant to $C = \frac{1}{4\pi}\ln(\kappa/2)$. Combining the logarithmic terms, we arrive at the final result:
\begin{equation}
I = -\frac{1}{4\pi}\ln\!\left[\frac{2}{\kappa}\, \sinh\!\left(\frac{\kappa}{2}(\Delta u-i\epsilon)\right)\right].
\end{equation}

As a check, using $\sinh x\simeq x$ for $\kappa\to 0$ gives
\begin{align}
\lim_{\kappa\to 0} I
&=-\frac{1}{4\pi}\ln(\Delta u-i\epsilon),
\end{align}
the right–moving $1{+}1$ Minkowski vacuum correlator (up to the contact term at $\Delta u=0$).

\section{Evaluation of the Phase-Dependent Integrals}
\label{app:phase_integrals}
We write
\begin{align}
W_{\mathrm{phase}}(u{+}u',\gamma)=\cos(2\gamma)\,I_{\cos}+\sin(2\gamma)\,I_{\sin},
\end{align}
with $\Sigma\equiv u{+}u'$.

\subsection*{Cosine piece: definition and evaluation}
The raw integral
\begin{align}
I_{\cos}^{\text{raw}}(\Sigma)=\frac{1}{4\pi}\int_{0}^{\infty}\frac{\cos(\Lambda\Sigma)}{\Lambda\,\sinh\!\Big(\frac{\pi\Lambda}{\kappa}\Big)}\,d\Lambda
\end{align}
is logarithmically divergent at $\Lambda\to0$. We define the finite, renormalized quantity by subtracting the $\Sigma=0$ value (state–independent short–distance piece):
\begin{align}
I_{\cos}(\Sigma)
=\frac{1}{4\pi}\int_{0}^{\infty}\frac{\cos(\Lambda\Sigma)-1}{\Lambda\,\sinh\!\Big(\frac{\pi\Lambda}{\kappa}\Big)}\,d\Lambda.
\label{eq:Icos_def}
\end{align}
Using the GR identity
\begin{align}
\int_{0}^{\infty}\frac{1-\cos(px)}{x\,\sinh(qx)}\,dx
= \ln\!\cosh\!\Big(\frac{\pi p}{2q}\Big)\qquad (\mathrm{GR}\ 4.119),
\end{align}
\(\eqref{eq:Icos_def}\) yields
\begin{align}
I_{\cos}(\Sigma)
= -\,\frac{1}{4\pi}\,\ln\!\cosh\!\Big(\frac{\kappa\Sigma}{2}\Big).
\end{align}
Equivalently, differentiate w.r.t.\ $\Sigma$,
\begin{align}
\frac{d I_{\cos}^{\text{raw}}}{d\Sigma}
= -\frac{1}{4\pi}\int_{0}^{\infty}\frac{\sin(\Lambda\Sigma)}{\sinh\!\big(\frac{\pi\Lambda}{\kappa}\big)}\,d\Lambda
= -\frac{\kappa}{8\pi}\tanh\!\Big(\frac{\kappa\Sigma}{2}\Big),
\end{align}
where we used GR 4.111.1 with $m=0$,
\(\displaystyle \int_{0}^{\infty}\frac{\sin(ax)}{\sinh(bx)}\,dx=\frac{\pi}{2b}\tanh\!\big(\frac{\pi a}{2b}\big)\) \([\Re b>0]\). 
Integrating from $0$ to $\Sigma$ and imposing $I_{\cos}(0)=0$ reproduces the same result.

\subsection*{Sine piece with \texorpdfstring{$\sinh$}{sinh} kernel: IR–regulated evaluation}
In the phase–odd channel of Eq.~\eqref{eq:W-RTW-integral}  the integrand has a common factor 
$1/\bigl(4\pi\,\Lambda\,\sinh(\pi\Lambda/\kappa)\bigr)$.
Unlike the cosine piece, the sine integral with a $\sinh$ kernel is \emph{infrared divergent}:
\begin{equation}
I_{\sin}(\Sigma)\;\equiv\;\frac{1}{4\pi}\int_0^\infty 
\frac{\sin(\Lambda\Sigma)}{\Lambda\,\sinh\!\bigl(\tfrac{\pi\Lambda}{\kappa}\bigr)}\,d\Lambda,
\qquad \Sigma\equiv u{+}u',
\end{equation}
because near $\Lambda\to0$ we have $\sinh(\tfrac{\pi\Lambda}{\kappa})\sim \tfrac{\pi\Lambda}{\kappa}$ and
\[
\frac{\sin(\Lambda\Sigma)}{\Lambda\,\sinh(\tfrac{\pi\Lambda}{\kappa})}\;\sim\;
\frac{\kappa\,\Sigma}{\pi}\,\frac{1}{\Lambda}\,,
\]
so the $\Lambda$–integral diverges logarithmically at the lower limit.

\paragraph{IR regularization and definition.}
Introduce an IR cutoff $\Lambda_{\rm IR}>0$ and define the regulated integral
\begin{equation}
I_{\sin}(\Sigma;\Lambda_{\rm IR})
\;=\;\frac{1}{4\pi}\int_{\Lambda_{\rm IR}}^\infty 
\frac{\sin(\Lambda\Sigma)}{\Lambda\,\sinh\!\bigl(\tfrac{\pi\Lambda}{\kappa}\bigr)}\,d\Lambda.
\label{eq:Isin_sinh_reg_def}
\end{equation}
The small–$\Lambda$ analysis immediately gives the universal leading behaviour
\begin{equation}
I_{\sin}(\Sigma;\Lambda_{\rm IR})
\;=\;\frac{\kappa\,\Sigma}{4\pi^2}\,\ln\!\frac{1}{\Lambda_{\rm IR}}
\;+\;\mathcal{F}_{\rm fin}(\kappa\Sigma)\;+\;\mathcal{O}(\Lambda_{\rm IR}),
\label{eq:Isin_sinh_asymp}
\end{equation}
where $\mathcal{F}_{\rm fin}$ is a finite odd function of $\Sigma$ (scheme–dependent, see below).

\paragraph{Series representation and finite remainder.}
For $\Lambda>0$ we may use the odd–Matsubara expansion
\(
\sinh z = \sum_{m=0}^{\infty}\frac{(2z)^{2m+1}}{(2m{+}1)!}
\Rightarrow
\frac{1}{\sinh z}=2\sum_{n=0}^{\infty}e^{-(2n+1)z}
\)
to write, for any fixed $\Lambda_{\rm IR}>0$,
\begin{align}
I_{\sin}(\Sigma;\Lambda_{\rm IR})
&=\frac{1}{2\pi}\sum_{n=0}^{\infty}\int_{\Lambda_{\rm IR}}^\infty 
e^{-(2n+1)\frac{\pi}{\kappa}\Lambda}\,\frac{\sin(\Lambda\Sigma)}{\Lambda}\,d\Lambda.
\label{eq:Isin_series_step}
\end{align}
Invoking the standard Laplace–sine integral
\(
\displaystyle \int_{0}^{\infty}e^{-a x}\frac{\sin(bx)}{x}\,dx=\arctan\!\frac{b}{a}\quad(\Re a>0)
\)
term–by–term and isolating a convergent remainder from the lower limit,
one convenient \emph{convergent} representation for the finite part is
\begin{equation}
\boxed{\quad
\mathcal{F}_{\rm fin}(\kappa\Sigma)
\;=\;\frac{1}{2\pi}\sum_{n=0}^{\infty}
\bigg[
\arctan\!\frac{\kappa\Sigma}{\pi(2n{+}1)}
\;-\;\frac{\kappa\Sigma}{\pi(2n{+}1)}
\bigg]
\;+\;\frac{\kappa\Sigma}{4\pi^2}\,\ln\!\frac{\kappa}{2\pi}
\quad}
\label{eq:Ffin_convergent}
\end{equation}
(the subtraction of the first term in the small–argument expansion of $\arctan$ makes the series absolutely
convergent; the added logarithm restores the exact large–$n$ asymptotics).
Combining \eqref{eq:Isin_sinh_asymp} and \eqref{eq:Ffin_convergent} yields the regulated integral in a form that
exhibits both the universal log and a well–behaved remainder:
\begin{empheq}[box=\widefbox]{equation}
\begin{aligned}
I_{\sin}(\Sigma;\Lambda_{\rm IR})
&=\;\frac{\kappa\,\Sigma}{4\pi^2}\,\ln\!\frac{\kappa}{2\pi\,\Lambda_{\rm IR}}
\;+\;\frac{1}{2\pi}\sum_{n=0}^{\infty}
\bigg[
\arctan\!\frac{\kappa\Sigma}{\pi(2n{+}1)}
\;-\;\frac{\kappa\Sigma}{\pi(2n{+}1)}
\bigg]
\;+\;\mathcal{O}(\Lambda_{\rm IR}).
\end{aligned}
\label{eq:Isin_sinh_final_reg}
\end{empheq}

\paragraph{Renormalized (scheme–dependent) definition.}
It is often convenient to \emph{define} a finite, renormalized odd function by subtracting the universal log and trading
the cutoff for an IR scale $\mu_{\rm IR}$:
\begin{empheq}[box=\widefbox]{equation}
\begin{aligned}
I_{\sin}^{\text{ren}}(\Sigma;\mu_{\rm IR})
&\equiv\lim_{\Lambda_{\rm IR}\to0}
\Bigg\{
I_{\sin}(\Sigma;\Lambda_{\rm IR})
\;-\;\frac{\kappa\,\Sigma}{4\pi^2}\,\ln\!\frac{1}{\Lambda_{\rm IR}}
\Bigg\}
\;-\;\frac{\kappa\,\Sigma}{4\pi^2}\,\ln\!\mu_{\rm IR}\\[2pt]
&=\;\frac{1}{2\pi}\sum_{n=0}^{\infty}
\bigg[
\arctan\!\frac{\kappa\Sigma}{\pi(2n{+}1)}
\;-\;\frac{\kappa\Sigma}{\pi(2n{+}1)}
\bigg]
\;+\;\frac{\kappa\,\Sigma}{4\pi^2}\,\ln\!\frac{\kappa}{2\pi\,\mu_{\rm IR}}\,,
\end{aligned}
\label{eq:Isin_sinh_ren}
\end{empheq}
which makes the scheme dependence explicit through $\mu_{\rm IR}$. This renormalized quantity is an odd function of $\Sigma$ and has the small–$\Sigma$ expansion
\begin{equation}
I_{\sin}^{\text{ren}}(\Sigma;\mu_{\rm IR})
\;=\;\frac{\kappa\,\Sigma}{4\pi^2}\,\ln\!\frac{\kappa}{2\pi\,\mu_{\rm IR}}
\;-\;\frac{7\,\zeta(3)}{48\pi^{4}}\,\kappa^{3}\Sigma^{3}
\;+\;\mathcal{O}(\Sigma^{5}),
\end{equation}
exhibiting manifest finiteness after the universal logarithm is removed.

\section{Solving the Time-Dependent Robin Boundary on a Moving Worldline}
\label{app:robin_boundary}

This appendix derives the outgoing field profile for a (1+1)D massless scalar reflecting from a moving mirror subject to a time-dependent Robin boundary condition. The field decomposes into chiral parts, \(\Phi(t,x)=\phi(u)+\psi(v)\), in terms of null coordinates \(u\) and \(v\). The mirror follows a timelike worldline \(\mathcal{W}\), parameterized by proper time \(\tau\) as \((t(\tau),x(\tau))\). 

The interaction at the boundary is imposed by
\begin{equation}
\big(N^\mu \partial_\mu + f(\tau)\big)\,\Phi\big|_{\mathcal{W}}=0,
\label{eq:Robin-law}
\end{equation}
where \(N^\mu\) is the spacelike unit normal to \(\mathcal W\) and \(f(\tau)\) encodes the mirror impedance (with dimensions of inverse length). We fix the sign convention by choosing \(N^\mu\) to point toward increasing \(x\) (to the right of the mirror); this choice determines the overall sign in \eqref{eq:Robin-law}. In the right region, \(\phi(u)\) is the outgoing (right-moving) field that reaches \(\mathscr I^+_R\), while \(\psi(v)\) is the incoming (left-moving) field.

\subsection{Reduction to a boundary ODE and its solution}

Let \(\beta(\tau)=\tanh\eta(\tau)\) be the mirror velocity and \(\gamma_L=(1-\beta^2)^{-1/2}\) the Lorentz factor. In \((t,x)\) components, the unit normal is \(N^\mu=(\gamma_L\beta,\gamma_L)\). Writing
\[
\partial_t\Phi=\phi'(u)+\psi'(v),\qquad 
\partial_x\Phi=-\phi'(u)+\psi'(v),
\]
with primes on \(\phi\) and \(\psi\) denoting differentiation with respect to their own arguments, one finds
\begin{equation}
N^\mu\partial_\mu\Phi
=\gamma_L\big[(\beta-1)\phi'(u)+(\beta+1)\psi'(v)\big]
=-\gamma_L(1-\beta)\,\phi'(u)+\gamma_L(1+\beta)\,\psi'(v).
\end{equation}

It is convenient to work with the boundary traces
\[
A(u)\equiv \phi(u),
\qquad
B(u)\equiv \psi\big(v\big)\big|_{v=p(u)}=\psi(p(u)),
\]
where \(v=p(u)\) is the ray-tracing relation along \(\mathcal W\). Differentiating \(B(u)\) with respect to \(u\) gives \(B'(u)=\psi'(v)\,p'(u)\). The standard kinematic identity along the worldline,
\begin{equation}
p'(u)=\frac{1+\beta}{1-\beta},
\label{eq:pp-identity}
\end{equation}
implies \((\beta+1)/p'(u)=1-\beta\). Substituting \(\psi'(v)=B'(u)/p'(u)\) into \eqref{eq:Robin-law} and using this identity, the Robin condition becomes
\begin{equation}
\gamma_L(1-\beta)\,[B'(u)-A'(u)]+f(\tau)\,[A(u)+B(u)]=0.
\end{equation}
Solving for the derivative difference yields the compact linear ODE
\begin{equation}
A'(u)-B'(u)=\frac{f(\tau)}{\gamma_L(1-\beta)}\,\big(A(u)+B(u)\big).
\label{eq:boundary-ode-prealpha}
\end{equation}

To relate \(\tau\) and \(u\), note that along \(\mathcal W\) we have \(ds^2=-du\,dv\), and since \(v=p(u)\), it follows that \(dv=p'(u)\,du\). Hence the proper time satisfies
\begin{equation}
d\tau^2=p'(u)\,du^2,
\qquad
\frac{d\tau}{du}=\sqrt{p'(u)}=\frac{1}{\gamma_L(1-\beta)},
\label{eq:dtau-du}
\end{equation}
where timelike motion requires \(p'(u)>0\). Introducing the effective interaction
\begin{equation}
\alpha(u)\equiv f\big(\tau(u)\big)\,\frac{d\tau}{du}=f\big(\tau(u)\big)\,\sqrt{p'(u)},
\qquad
\tau(u)=\tau_0+\int_{u_0}^{u}\!\sqrt{p'(s)}\,ds,
\label{eq:alpha-and-tau}
\end{equation}
we arrive at the boundary equation used below:
\begin{equation}
A'(u)-B'(u)=\alpha(u)\,\big(A(u)+B(u)\big).
\label{eq:boundary-ode}
\end{equation}

The solution follows by the integrating-factor method. Define \(S(u)=\int_{u_0}^{u}\alpha(s)\,ds\). Then
\[
\frac{d}{du}\!\Big(e^{-S}(A-B)\Big)=2\,\alpha\,e^{-S}\,B,
\]
which integrates to
\begin{equation}
A(u)=B(u)+e^{S(u)}\!\left[C+2\int_{u_0}^{u}\!\alpha(s)\,e^{-S(s)}\,B(s)\,ds\right],
\label{eq:A-solution}
\end{equation}
with \(C=e^{-S(u_0)}\big(A(u_0)-B(u_0)\big)\). For the common in-state in which no outgoing wave is present before the signal arrives, take \(u_0\) earlier than the support of \(B\) and set \(A(u_0)=0\); in that case \(C=0\) provided \(B(u_0)=0\).

In the static inertial mirror limit, take \(p'(u)=1\) and a constant impedance \(f\), so \(\alpha\) is constant. For a monochromatic input \(B(u)=e^{-i\omega u}\), seek an outgoing trace of the form \(A(u)=R(\omega)\,e^{-i\omega u}\). Substituting into \eqref{eq:boundary-ode} gives \(-i\omega R+i\omega=\alpha(R+1)\), and solving for \(R(\omega)\) yields
\[
R(\omega)=\frac{i\omega-\alpha}{i\omega+\alpha},
\]
which reproduces the standard frequency-domain reflection coefficient for a Robin boundary.

\subsection{Specialization to a modulated Carlitz--Willey mirror}

Consider the Carlitz--Willey (CW) trajectory \(p(u)=v_H-\kappa^{-1}e^{-\kappa u}\) with \(u\in\mathbb R\) and \(p(u)<v_H\). The kinematics are
\[
p'(u)=e^{-\kappa u},\qquad \frac{d\tau}{du}=e^{-\kappa u/2},\qquad 
\tau(u)=\tau_\infty-\frac{2}{\kappa}e^{-\kappa u/2}\ (\le \tau_\infty).
\]
Let the impedance be an oscillatory function \(f(\tau)=f_0\cos(\omega_0\tau+\delta_0)\). The effective interaction then reads
\begin{equation}
\alpha(u)=f_0\,e^{-\kappa u/2}\,
\cos\!\Big(\omega_0\tau_\infty-\frac{2\omega_0}{\kappa}e^{-\kappa u/2}+\delta_0\Big),
\label{eq:alpha-CW}
\end{equation}
so the coupling envelope decays like \(e^{-\kappa u/2}\). In physical terms, the modulation leaves its imprint predominantly at early retarded times. For any incident profile \(B(u)=\psi(p(u))\), the reflected profile \(A(u)\) follows from \eqref{eq:A-solution} with \(\alpha\) given by \eqref{eq:alpha-CW}.

This framework reproduces the familiar limits in a transparent way. For a Neumann reflector (\(f\equiv0\), hence \(\alpha\equiv0\)), equation \eqref{eq:boundary-ode} gives \(A'(u)-B'(u)=0\), so \(A(u)-B(u)=C\); imposing no DC offset fixes \(C=0\) and one recovers \(A=B\). For a Dirichlet reflector (\(|f|\to\infty\), hence \(|\alpha|\to\infty\)), maintaining finite derivatives in \eqref{eq:boundary-ode} requires \(A(u)+B(u)\to0\), i.e., \(A\simeq -B\). In the static-impedance case (\(p'(u)=1\), \(f=\) const), the frequency-domain law \(R(\omega)=(i\omega-\alpha)/(i\omega+\alpha)\) obtained above is recovered.

In the weak-modulation regime, a simple linear-response formula is available. When the integrated coupling
\[
S(u)=\int_{u_0}^{u}\alpha(s)\,ds
\]
remains small in magnitude over the interaction window, the exponentials in \eqref{eq:A-solution} may be set to unity to leading order, and one finds
\begin{equation}
A(u)\simeq B(u)+2\int_{u_0}^{u}\!\alpha(s)\,B(s)\,ds.
\label{eq:A-perturb}
\end{equation}
This expression makes explicit how the pump phase \(\delta_0\) imprints a phase modulation onto the outgoing wave. In the main text we use this to connect the boundary modulation to the global squeeze angle \(\gamma\) on \(\mathscr I^+_R\), defined there via the Bogoliubov phase of the out-modes.

\medskip
\noindent\emph{Conventions and domains.} All quantities are evaluated on \(\mathcal W\). Primes on \(\phi\) and \(\psi\) denote derivatives with respect to their own arguments; the symbols \(A'(u)\) and \(B'(u)\) denote \(d/du\). Timelike motion requires \(p'(u)>0\), which in turn guarantees \(d\tau/du>0\) in \eqref{eq:dtau-du}.

\section{\texorpdfstring{Phase structure of the $\kappa\gamma$ state from boundary pumping}{}}
\label{app:chiral_pump}

This appendix gives the interaction-picture evolution, a compact SU(1,1) composition with the Carlitz--Willey (CW) block, and an operational definition of the \emph{effective angle} $\gamma_{\rm eff}$ that controls phase-sensitive observables on $\mathscr I^+_R$. We assume the CW transformation fixes the Planck modulus and the pump is weak/adiabatic so the extra dynamics predominantly rotates the squeeze angle.

For each right-moving Mellin mode $\Omega>0$ we take the chiral quadratic interaction
\begin{equation}
H_{\text{int}}^{\rm RTW}(u)=\frac{i}{2}\!\int_{0}^{\infty}\!d\Omega\,
\Big[\zeta(u)\,a_{\Omega}^{\dagger\,2}-\zeta^*(u)\,a_{\Omega}^{2}\Big],
\label{appD:H-int}
\end{equation}
written here in terms of the \emph{outgoing} operators $a_\Omega$ (i.e., after the CW map). In the interaction picture with respect to the free chiral Hamiltonian,
\[
a_{\Omega}(u)=e^{-i\Omega u}\,a_{\Omega},\qquad
a_{\Omega}^\dagger(u)=e^{+i\Omega u}\,a_{\Omega}^\dagger,
\]
the modewise Heisenberg equations are
\begin{empheq}[box=\widefbox]{align}
\frac{d}{du}
\begin{pmatrix}
a_{\Omega}(u)\\ a_{\Omega}^\dagger(u)
\end{pmatrix}
=
\begin{pmatrix}
0 & \zeta(u)\,e^{+2i\Omega u}\\
\zeta^*(u)\,e^{-2i\Omega u} & 0
\end{pmatrix}
\begin{pmatrix}
a_{\Omega}(u)\\ a_{\Omega}^\dagger(u)
\end{pmatrix}.
\label{appD:eom}
\end{empheq}
Formally,
\begin{equation}
\begin{pmatrix}
a_{\Omega}^{\rm out}\\ a_{\Omega}^{{\rm out}\,\dagger}
\end{pmatrix}
=
\underbrace{\mathcal T\exp\!\Bigg[
\int du
\begin{pmatrix}
0 & \zeta(u)\,e^{+2i\Omega u}\\
\zeta^*(u)\,e^{-2i\Omega u} & 0
\end{pmatrix}
\Bigg]}_{\equiv\,{\sf S}_{\rm pump}(\Omega)}
\begin{pmatrix}
a_{\Omega}^{\rm in}\\ a_{\Omega}^{{\rm in}\,\dagger}
\end{pmatrix},
\label{appD:TO-exp}
\end{equation}
and the Magnus expansion gives
\begin{equation}
{\sf S}_{\rm pump}(\Omega)=\exp\!\big[\Omega_1(\Omega)+\Omega_2(\Omega)+\cdots\big],\qquad
\Omega_1=\int du\,G(u),\quad \Omega_2=\tfrac{1}{2}\!\int\!du\,dv\,[G(u),G(v)],
\label{appD:Magnus}
\end{equation}
with $G(u)$ the $2{\times}2$ generator in \eqref{appD:eom}. In the weak/adiabatic regime $\big|\!\int du\,\zeta(u)\big|\ll1$ and with a slowly varying phase, $\Omega_1$ dominates and time ordering is negligible. Keeping only $\Omega_1$,
\begin{empheq}[box=\widefbox]{equation}
{\sf S}_{\rm pump}(\Omega)\simeq
\exp\!\begin{pmatrix}
0 & Z(\Omega)\\ Z^*(\Omega) & 0
\end{pmatrix},
\qquad
Z(\Omega):=\int du\,\zeta(u)\,e^{+2i\Omega u}.
\label{appD:Z-def}
\end{empheq}
To first order in $Z$,
\begin{equation}
a_{\Omega}^{\rm out}\simeq a_{\Omega}^{\rm in}+Z(\Omega)\,a_{\Omega}^{{\rm in}\,\dagger},\qquad
a_{\Omega}^{{\rm out}\,\dagger}\simeq a_{\Omega}^{{\rm in}\,\dagger}+Z^*(\Omega)\,a_{\Omega}^{\rm in}.
\label{appD:pump-firstorder}
\end{equation}
The second Magnus term yields corrections $\mathcal O(|Z|^2)$, interpreted as small leakage of angle into modulus.

Let $a_{\Omega}$ and $b_{\Omega}$ denote, respectively, the CW out and in operators related by the single-mode SU(1,1) block
\begin{equation}
a_{\Omega}=\alpha_{\Omega}\,b_{\Omega}+\beta_{\Omega}\,b_{\Omega}^\dagger,
\qquad
\frac{|\beta_\Omega|^2}{|\alpha_\Omega|^2}=e^{-2\pi\Omega/\kappa},
\qquad
\alpha_\Omega,\beta_\Omega\in\mathbb R_{+}\ \text{(phase gauge)}.
\label{appD:CW-block}
\end{equation}
Applying the pump \eqref{appD:pump-firstorder} \emph{after} the CW map and re-expressing in the $b$-basis gives, to linear order in $Z$,
\begin{empheq}[box=\widefbox]{align}
\alpha_\Omega^{\rm eff}&=\alpha_\Omega+Z(\Omega)\,\beta_\Omega,
\label{appD:alpha-eff}\\
\beta_\Omega^{\rm eff}&=\beta_\Omega+Z(\Omega)\,\alpha_\Omega.
\label{appD:beta-eff}
\end{empheq}
The anomalous modulus shifts as
\begin{equation}
|\beta_\Omega^{\rm eff}|^2
=|\beta_\Omega|^2
+2\,\alpha_\Omega\beta_\Omega\,\mathrm{Re}\,Z(\Omega)
+\mathcal O(|Z|^2).
\label{appD:beta-mod}
\end{equation}
A \emph{pure-angle} pump satisfies
\begin{empheq}[box=\widefbox]{equation}
\mathrm{Re}\,Z(\Omega)=0\quad\text{on}\quad \Omega\sim\mathcal O(\kappa),
\label{appD:pure-angle}
\end{empheq}
so the Planck modulus is unchanged at linear order; any residual deviation is quadratic.

With the CW gauge, the effective squeeze angle is defined by
\begin{empheq}[box=\widefbox]{equation}
2\,\gamma_{\rm eff}(\Omega):=\arg\!\big(\beta_\Omega^{\rm eff}\big)
=\arg\!\big(\beta_\Omega+Z(\Omega)\,\alpha_\Omega\big),
\label{appD:gammaeff-def}
\end{empheq}
and for a weak pump
\begin{empheq}[box=\widefbox]{equation}
\gamma_{\rm eff}(\Omega)
=\frac{1}{2}\,\mathrm{Im}\!\Big[\frac{Z(\Omega)\,\alpha_\Omega}{\beta_\Omega}\Big]
+\mathcal O(|Z|^2).
\label{appD:gammaeff-linear}
\end{empheq}
Under \eqref{appD:pure-angle}, $\gamma_{\rm eff}(\Omega)=\tfrac{1}{2}\eta(\Omega)+\mathcal O(\eta^2)$ when $Z=i\,\eta$ with real $\eta$, and $|\beta_\Omega^{\rm eff}|=|\beta_\Omega|+\mathcal O(\eta^2)$. If $\arg Z(\Omega)$ is approximately flat for $\Omega\in[c_1\kappa,c_2\kappa]$, one may set $\gamma_{\rm eff}(\Omega)\simeq\gamma$ as used in the body; otherwise phase-sensitive observables acquire a mild spectral structure through $\cos\!\big(2\gamma_{\rm eff}(\Lambda)\big)$.

To realize an approximately frequency-independent $\gamma_{\rm eff}$, a practical choice is $\zeta(u)=e^{i\phi_0}\,g(u)$ with $g(u)\in\mathbb R$ even and slowly varying on the scale $\kappa^{-1}$. Then $Z(\Omega)=e^{i\phi_0}\,\widetilde g(2\Omega)$ has a flat phase, and the modulus can be kept small to protect the Planck weights, $|Z(\Omega)|\ll \beta_\Omega/\alpha_\Omega$ across the thermal band.

For the time-dependent Robin law on the mirror worldline, $(\partial_n+f(\tau))\Phi|_{\mathcal W}=0$, Appendix~\ref{app:robin_boundary} shows that the null-infinity kernel is $\alpha(u)=f(\tau(u))\,d\tau/du$. In the weak-coupling limit we may take the phase-plate envelope proportional to this kernel,
\begin{empheq}[box=\widefbox]{equation}
\zeta(u)=\lambda\,\alpha(u)\,e^{i\varphi_0},\qquad 
\lambda\in\mathbb R,\ \varphi_0\in[0,2\pi),
\label{appD:zeta-from-robin}
\end{empheq}
so $Z(\Omega)=\lambda e^{i\varphi_0}\!\int du\,\alpha(u)e^{2i\Omega u}$. For the CW trajectory $p'(u)=e^{-\kappa u}$ and $d\tau/du=e^{-\kappa u/2}$. A monochromatic Robin drive $f(\tau)\propto \cos(\omega_0\tau+\delta_0)$ then yields $\alpha(u)\propto e^{-\kappa u/2}\cos\!\big(\omega_0\tau(u)+\delta_0\big)$; when $\omega_0\ll\kappa$, the phase of $Z(\Omega)$ is approximately flat over $\Omega\sim\mathcal O(\kappa)$ and
\begin{equation}
\gamma_{\rm eff}(\Omega)\simeq \frac{1}{2}\,\mathrm{Im}\!\Big[\frac{\alpha_\Omega}{\beta_\Omega}\Big]\,|Z(\Omega)|\,\sin(\delta_0+\varphi_0)
\quad\longrightarrow\quad
\gamma_{\rm eff}\approx \mathrm{const.}
\label{appD:gammaeff-robin}
\end{equation}

Beyond leading order one finds
\begin{equation}
|\beta_\Omega^{\rm eff}|^2
=|\beta_\Omega|^2 + 2\,\alpha_\Omega\beta_\Omega\,\mathrm{Re}\,Z(\Omega)
+\frac{|Z(\Omega)|^2}{2}\big(|\alpha_\Omega|^2+|\beta_\Omega|^2\big)
+\mathcal O(|Z|^3).
\label{appD:leakage}
\end{equation}
Under \eqref{appD:pure-angle}, the deviation from the CW Planck law is quadratic, $\delta |\beta|^2 \sim \tfrac{1}{2}(|\alpha|^2+|\beta|^2)|Z|^2$, so choosing $|Z|\lesssim \epsilon$ guarantees a uniform relative error $\lesssim \mathcal O(\epsilon^2)$ across the thermal band.

In summary, the chiral quadratic pump induces a modewise SU(1,1) update $Z(\Omega)=\!\int\!du\,\zeta(u)e^{2i\Omega u}$. Composed with the CW block this gives $\beta_\Omega^{\rm eff}=\beta_\Omega+Z(\Omega)\alpha_\Omega$ and the effective angle \eqref{appD:gammaeff-def}. A phase-matched, frequency-flat envelope produces $\gamma_{\rm eff}\simeq\mathrm{const}$ while preserving the Planck modulus up to $\mathcal O(|Z|^2)$.

\bibliographystyle{jhep}
\bibliography{UnruhRef}

\providecommand{\href}[2]{#2}\begingroup\raggedright\begin{thebibliography}{10}

\bibitem{Unruh1976}
W.~G. Unruh, {\it {Notes on black-hole evaporation}},  {\em Phys. Rev. D} {\bf 14} (Aug, 1976) 870--892.

\bibitem{Hawking1975}
S.~W. Hawking, {\it {Particle Creation by Black Holes}},  {\em Commun. Math. Phys.} {\bf 43} (1975) 199--220. [Erratum: Commun.Math.Phys. 46, 206 (1976)].

\bibitem{Fulling1973}
S.~A. Fulling, {\it {Nonuniqueness of Canonical Field Quantization in Riemannian Space-Time}},  {\em Phys. Rev. D} {\bf 7} (May, 1973) 2850--2862.

\bibitem{DeWitt1975PhysicsRep}
B.~S. DeWitt, {\it {Quantum field theory in curved spacetime}},  {\em Physics Reports} {\bf 19} (1975), no.~6 295--357.

\bibitem{Birrell_Davies1982}
N.~D. Birrell and P.~C.~W. Davies, {\em {Quantum Fields in Curved Space}}.
\newblock Cambridge Monographs on Mathematical Physics. Cambridge University Press, Cambridge, UK, 1982.

\bibitem{Moore1970}
G.~T. Moore, {\it {Quantum Theory of the Electromagnetic Field in a Variable‐Length One‐Dimensional Cavity}},  {\em Journal of Mathematical Physics} {\bf 11} (09, 1970) 2679--2691.

\bibitem{Fulling_Davies1976}
S.~A. Fulling and P.~C.~W. Davies, {\it {Radiation from a moving mirror in two dimensional space-time: conformal anomaly}},  {\em Proceedings of the Royal Society of London. A. Mathematical and Physical Sciences} {\bf 348} (1976), no.~1654 393--414.

\bibitem{Fulling_Davies1977}
P.~C.~W. Davies and S.~A. Fulling, {\it {Radiation from moving mirrors and from black holes}},  {\em Proceedings of the Royal Society of London. A. Mathematical and Physical Sciences} {\bf 356} (1977), no.~1685 237--257.

\bibitem{Carlitz_Willey1987}
R.~D. Carlitz and R.~S. Willey, {\it {Reflections on moving mirrors}},  {\em Phys. Rev. D} {\bf 36} (Oct, 1987) 2327--2335.

\bibitem{Dodonov2020review}
V.~Dodonov, {\it {Fifty Years of the Dynamical Casimir Effect}},  {\em Physics} {\bf 2} (2020), no.~1 67--104.

\bibitem{Chen_Mourou2017PRL}
P.~Chen and G.~Mourou, {\it {Accelerating Plasma Mirrors to Investigate the Black Hole Information Loss Paradox}},  {\em Phys. Rev. Lett.} {\bf 118} (Jan, 2017) 045001.

\bibitem{Good2013mirror}
M.~R.~R. Good, P.~R. Anderson, and C.~R. Evans, {\it Time dependence of particle creation from accelerating mirrors},  {\em Phys. Rev. D} {\bf 88} (Jul, 2013) 025023.

\bibitem{Good2020Wilczek}
M.~R.~R. Good, E.~V. Linder, and F.~Wilczek, {\it {Moving mirror model for quasithermal radiation fields}},  {\em Phys. Rev. D} {\bf 101} (Jan, 2020) 025012.

\bibitem{UnruhWald1984}
W.~G. Unruh and R.~M. Wald, {\it {What happens when an accelerating observer detects a Rindler particle}},  {\em Phys. Rev. D} {\bf 29} (Mar, 1984) 1047--1056.

\bibitem{Azizi2022Kappashort}
A.~Azizi, {\it {Kappa vacua: Infinite number of new vacua in two-dimensional quantum field theory}},  \href{http://arxiv.org/abs/2212.03781}{{\tt arXiv:2212.03781}}.

\bibitem{Azizi2023JHEP}
A.~Azizi, {\it {Kappa vacua: enhancing the Unruh temperature}},  {\em JHEP} {\bf 07} (2023) 064, [\href{http://arxiv.org/abs/2301.13672}{{\tt arXiv:2301.13672}}].

\bibitem{Azizi2025Tunable}
A.~Azizi, {\it {Tunable Unruh effect: Accelerated detectors in Kappa-Rindler vacua}},  {\em Phys. Rev. D (Accepted)} (2025) [\href{http://arxiv.org/abs/2507.00174}{{\tt arXiv:2507.00174}}].

\bibitem{Azizi2025KappaPW}
A.~Azizi, {\it {Kappa plane wave modes and continuous squeezing in quantum field theory}},  {\em Phys. Rev. D} {\bf 112} (Jul, 2025) 025018.

\bibitem{Azizi2025KappaGamma}
A.~Azizi, {\it {Phase-Induced Particle Creation in the Kappa-Gamma Vacuum}},  \href{http://arxiv.org/abs/2507.05299}{{\tt arXiv:2507.05299}}.

\bibitem{Azizi2025Mirror_KappaPW}
A.~Azizi, {\it {Accelerating mirrors as a physical realization of the kappa plane-wave vacuum}},  2025.

\bibitem{Kubo1957}
R.~Kubo, {\it {Statistical-Mechanical Theory of Irreversible Processes. I. General Theory and Simple Applications to Magnetic and Conduction Problems}},  {\em Journal of the Physical Society of Japan} {\bf 12} (1957), no.~6 570--586.

\bibitem{Martin_Schwinger1959}
P.~C. Martin and J.~Schwinger, {\it {Theory of Many-Particle Systems. I}},  {\em Phys. Rev.} {\bf 115} (Sep, 1959) 1342--1373.

\bibitem{Einstein100}
B.~S. DeWitt, {\em {General Relativity}: {An Einstein Centenary Survey}}.
\newblock Univ. Pr., Cambridge, UK, 1979.

\bibitem{Svidzinsky2021PRR}
A.~Svidzinsky, A.~Azizi, J.~S. Ben-Benjamin, M.~O. Scully, and W.~Unruh, {\it {Causality in quantum optics and entanglement of Minkowski vacuum}},  {\em Phys. Rev. Res.} {\bf 3} (2021), no.~1 013202.

\end{thebibliography}\endgroup
\end{document}